\documentclass{PoS}
\usepackage[numbers, square, comma, sort&compress]{natbib}
\usepackage{amsmath}
\def\Tsu{\frac{{\bf T}_{s}^2-{\bf T}_{u}^2}{2}}
\def\Tt{{\bf T}_t^2}
\title{Recent progress on infrared singularities\thanks{Preprint number: Edinburgh 2018/2.}}

\ShortTitle{Recent progress on infrared singularities}

\author{\speaker{Einan Gardi}
     
\\
      Higgs Centre for Theoretical Physics, 
School of Physics and Astronomy, \\
The University of Edinburgh, Edinburgh EH9 3FD, Scotland, UK\\
      E-mail: \email{einan.gardi@ed.ac.uk}}

\abstract{Over the past couple of years we have had significant progress in determining long-distance singularities in gauge-theory scattering amplitudes of massless particles beyond the planar limit. 
Upon considering all kinematic invariants much larger than the QCD scale, the singularities factorise into universal soft and jet functions, leaving behind a finite hard-interaction amplitude. Such factorization can now be implemented in full to three loops for arbitrary scattering processes of massless partons. In particular, the soft anomalous dimension for a general configuration of $n$ coloured particles was computed to this order, where it displays for the first time non-dipole interactions that correlate the colour and kinematic degrees of freedom of three and four particles.  
In parallel, there has been progress in understanding amplitudes and their singularities in special kinematic limits, such as collinear limits of multi-leg amplitudes and the high-energy limit in $2\to 2$ scattering. These relate respectively to different factorization properties of gauge-theory amplitudes. In this talk I describe the state of the art and illustrate the interplay between the analysis of the singularities for general kinematics and the properties of amplitudes in special kinematic limits. 
}

\FullConference{13th International Symposium on Radiative Corrections (Applications of Quantum Field Theory to Phenomenology)\\
		 25-29 September, 2017 \\
		 St. Gilgen, Austria}

\begin{document}

\section{Introduction}

The study of gauge-theory scattering amplitudes at the multi-loop level is central to precision collider physics, as well as to the fundamental understanding of these theories. 
A salient property of gauge-theory scattering amplitudes is the presence of infrared singularities. Understanding the origin and structure of these singularities is essential for cross-section calculations, where one of the biggest challenges is the intricate cancellation of the singularities between virtual and real corrections. For sufficiently inclusive quantities, this knowledge can be translated into all-order resummation of various towers of logarithms.
From a more theoretical perspective, understanding the singularity structure of amplitudes is interesting since it provides deep insight into all-order properties of amplitudes and their iterative structure. 
Indeed, one of the main features of infrared singularities is that they exponentiate, and can thus be encoded into an anomalous dimension. The latter can be computed from the ultraviolet singularities of correlators of semi-infinite Wilson line
operators~\cite{Polyakov:1980ca,Arefeva:1980zd,Dotsenko:1979wb,Brandt:1981kf,Korchemsky:1985xj,Korchemsky:1985xu,Korchemsky:1987wg}. 
Infrared singularities are therefore simpler than finite corrections, and they feature a transparent iterative structure which is similar across different gauge theories. 

Further insight into the all-order structure of scattering amplitudes may be gained by considering special kinematic limits, where new factorization properties arise. 
The simplification owing to special kinematic limits may render both infrared-divergent and finite terms computable to high loop orders, and potentially even to all orders.
For example, the high-energy (Regge) limit of two-parton scattering (forward scattering with $s\gg -t$) lends itself to an effective description in terms infinite Wilson lines~\cite{Korchemskaya:1996je,Korchemskaya:1994qp}, or alternatively in terms of Reggeized gluons propagating in transverse space, i.e. in $2-2\epsilon$ dimensions. The wavefunction describing Reggeised gluons admits a rapidity evolution equation, the Balitsky-Fadin-Kuraev-Lipatov (BFKL) equation and its generalizations~\cite{Kuraev:1977fs,Balitsky:1978ic,Lipatov:1985uk,Mueller:1993rr,Mueller:1994jq,Balitsky:1995ub,Balitsky:1998kc,Kovchegov:1999yj,JalilianMarian:1996xn,JalilianMarian:1997gr,Iancu:2001ad},  making it possible to sum up high-energy logarithms to all orders~\cite{Caron-Huot:2013fea,Caron-Huot:2017fxr,Caron-Huot:2017zfo}.
Another example where major simplifications occur are collinear limits of multileg amplitudes. Specifically, one may identify a relation between an $n$-parton amplitude in which two partons are collinear and an $(n-1)$-parton amplitude, where the collinear divergences in the former are factorized into a \emph{splitting amplitude}. The universality of the splitting amplitude translates into constraints on both the finite and infrared-divergent parts of amplitudes.

Infrared singularities have been studied in QCD for many decades for processes
involving two partons and any number of colour singlet
particles~\cite{Mueller:1979ih,Collins:1980ih,Sen:1981sd,Sen:1982bt,Gatheral:1983cz,Frenkel:1984pz,Sterman:1981jc,Magnea:1990zb}. The
singularity structure of amplitudes involving several partons has been
examined more recently, in both the
massless~\cite{Korchemsky:1993hr,Korchemskaya:1996je,Korchemskaya:1994qp,Catani:1996vz,Catani:1998bh,Sterman:2002qn,Dixon:2008gr,Kidonakis:1998nf,Bonciani:2003nt,Dokshitzer:2005ig,Aybat:2006mz,Gardi:2009qi,Becher:2009cu,Becher:2009qa,Gardi:2009zv,
  Dixon:2009gx,Dixon:2009ur,Bret:2011xm,DelDuca:2011ae,Caron-Huot:2013fea,Ahrens:2012qz,Naculich:2013xa,Erdogan:2014gha,Gehrmann:2010ue}
and
massive~\cite{Kidonakis:2009ev,Mitov:2009sv,Becher:2009kw,Beneke:2009rj,Czakon:2009zw,Ferroglia:2009ii,Chiu:2009mg,Mitov:2010rp,Mitov:2010xw,Gardi:2013saa,Falcioni:2014pka,Henn:2013fah,Dukes:2013gea,Dukes:2013wa,Gardi:2010rn,Gardi:2011wa,Gardi:2011yz,Gardi:2013ita,Laenen:2008gt,Vladimirov:2015fea}
cases.   
In ref.~\cite{Almelid:2015jia} the multileg soft anomalous dimension for massless particles has been determined at three-loop order. This diagrammatic  
calculation was reviewed  in ref.~\cite{Gardi:2016ttq}. 
The present talk concerns primarily the properties of three-loop soft anomalous dimension following ref.~\cite{Almelid:2017qju}.
We begin by briefly reviewing the factorization and exponentiation properties of soft singularities, followed by a description of the general structure of the soft anomalous dimension.  Next, rather than analysing the three-loop result directly, we will present
general considerations allowing one to identify the relevant kinematic variables and the class of iterated integrals in terms of which the result can be written. Subsequently, we will examine the two kinematic limits mentioned above, where information regarding soft singularities can be deduced using other methods. Considering the high-energy limit we will summarise recent conclusions  regarding soft singularities based on rapidity evolution equations according to~Refs.~\cite{Caron-Huot:2017fxr,Caron-Huot:2017zfo}.
Finally, we will discuss the bootstrap procedure of ref.~\cite{Almelid:2017qju}, which  showed that, quite remarkably, it is possible to start with a completely general
ansatz of basis functions, and recover the three-loop result of~\cite{Almelid:2015jia}, up to an overall normalization factor, by imposing symmetry considerations along with properties of the singularities in the collinear and high-energy limits.

\section{Factorization and exponentiation of soft singularities}

The most important fact about infrared singularities is that they can be factorized: 
the singularities emerge from soft or collinear modes, which do not resolve many details of the process associated with highly virtual particles, and therefore factorize.
Our interest in this talk is in scattering of \emph{massless} quarks and gluons, which means that there are collinear singularities, in addition to soft ones. 
In \emph{fixed-angle factorization} we consider all the hard scales in the problem, $s_{ij}$, to be much larger than the QCD scale $\Lambda^2$: 
\[
s_{ij}=2p_i\cdot p_j=2\beta_i\cdot \beta_j \,Q^2\gg \Lambda^2\,,
\] 
where the 4-velocity vector $\beta_i$ is proportional to the momentum $p_i$, having extracted an overall hard scale factor $Q$.
Infrared singularities emerge from soft and collinear regions, which are captured by separate soft and jet functions. The amplitude for $n$ parton scattering in $d=4-2\epsilon$
spacetime dimensions  assumes the factorised
form~\cite{Sen:1982bt,Kidonakis:1998nf,Sterman:2002qn,Dixon:2008gr,Aybat:2006mz,Gardi:2009qi,Gardi:2009zv}
\begin{equation}
{\cal A}_n(\{p_i\},\epsilon,\alpha_s)=
{\cal S}(\{\beta_i\},\{{\bf T}_i \},\epsilon,\alpha_s)\,
{\cal H}_n(\{p_i\},\{n_i\},\epsilon,
\alpha_s)\,
\prod_{i=1}^n \frac{J(p_i,n_i,\epsilon,\alpha_s)}
{{\cal J}(\beta_i,n_i,\epsilon,\alpha_s)},
\label{softcolfac}
\end{equation}
where dependence on colour degrees of freedom is shown in terms of colour operators ${\bf T}_i$ for each particle $i$, $\alpha_s$ is the $d$-dimensional running coupling, ${\cal H}_n$ is a process-dependent {\it hard function} that is finite as
$\epsilon\rightarrow 0$, and ${\cal S}(\{\beta_i\},\{{\bf T}_i \},\epsilon,\alpha_s)$ and $J(p_i,n_i,\epsilon,\alpha_s)$ are the {\it soft} and {\it
  jet functions} that collect infrared singularities originating from
emissions that are soft and collinear to particle $i$
respectively. The function ${\cal J}(\beta_i,n_i,\epsilon,\alpha_s)$ in (\ref{softcolfac}) removes the double counting of soft-collinear divergences accounted for in both ${\cal S}$ and $J$.
 
For our discussion the soft function ${\cal S}$ will be most important, since only this function involves singularities depending on the colour and kinematic degrees of freedom of several partons, and as such becomes significantly more complicated in multileg scattering beyond the planar limit as compared to two-parton scattering, or indeed the planar limit in multileg scattering.
The soft function can be defined and computed using a correlator of products of semi-infinite Wilson lines, emanating from the point of the hard interaction:
\begin{equation}
{\cal S}(\{\beta_i\},\{{\bf T}_i\} , \epsilon,\alpha_s)=\left\langle 0\left| 
{\rm T}\left[\Phi_{\beta_1}\,\Phi_{\beta_2}\ldots\Phi_{\beta_n}\right]
\right|0\right\rangle,
\label{Sdef}
\end{equation}
where ${\rm T}[\cdots]$ represents a time-ordered product and 
\begin{equation}
\Phi_{\beta_i}={\cal P}\exp\left[i\mu^\epsilon g_s {\bf
    T}_i^a\int_0^\infty ds\,\beta_i\cdot A^a(s\beta_i)\right]
\label{Phidef}
\end{equation}
is a Wilson-line operator along a trajectory of the $i^{\rm th}$
parton.
The soft function therefore depends on the angles between the lines. It also has a non-trivial colour structure which acts as a matrix in colour space on the hard interaction.  The jets, in turn, are  colour singlet and hence simple: collinear singularities do not depend on the colour flow of the other partons and they depend on the momentum of a single parton.

The next essential property of infrared singularities is that they exponentiate. This can be shown most easily by establishing that the soft function (and similarly the jet function) admits a renormalization-group equation, whose solution is a path-ordered exponential. Upon combining the singular factors in eq.~(\ref{softcolfac}), the factorized amplitude may be expressed as~\cite{Becher:2009qa,Gardi:2009zv,Gardi:2016ttq}
\begin{equation}
{\cal
  A}_n\left(\{p_i\},\epsilon,\alpha_s(\mu^2)
\right)=
Z_n\left(\{p_i\},\{{\bf T}_i\} ,\epsilon,\alpha_s(\mu_f^2)
\right)
{\cal H}_n\left(\{p_i\},\frac{\mu_f}{\mu}
,\epsilon,\alpha_s(\mu^2)\right)\,,
\label{Andef}
\end{equation}
where the factor $Z_n$, which captures all the singularities, is a matrix in colour-flow space, and can be written as
\begin{equation}
Z_n\left(\{p_i\},\{{\bf T}_i\} ,\epsilon,\alpha_s(\mu_f^2)
\right)={\cal
  P}\exp\left\{-\frac12\int_0^{\mu_f^2}\frac{d\lambda^2}{\lambda^2}
\Gamma_n\left(\{p_i\},\{{\bf T}_i \},\lambda,
\alpha_s(\lambda^2)\right)\right\},
\label{ZnRG}
\end{equation}
where $\mu_f^2$ is a factorization scale, ${\cal P}$ denotes path-ordering and $\Gamma_n$ is the so-called \emph{soft anomalous dimension}. The latter encodes all infrared singularities, order by order in the perturbative expansion. It is itself a finite quantity, and the singularities in eq.~(\ref{ZnRG}) are generated by integrating over the $d$-dimensional coupling from zero momentum. In the rest of the talk I will focus on the kinematics and colour dependence of the soft anomalous dimension $\Gamma_n$.

\section{The structure of the soft anomalous dimension}

At one loop a single gluon may be exchanged between any of the Wilson lines in (\ref{Sdef}), leading to a very simple structure in  $\Gamma_n$: it is just a sum over colour dipoles, ${\bf T}_i\cdot {\bf T}_j$ in colour space, multiplied by the logarithm of the corresponding kinematic invariant $s_{ij}$. Remarkably, it turns out that this dipole sum formula generalises to two-loops. This was first observed in an explicit computation~\cite{Aybat:2006mz} demonstrating 
that potential three-particle correlations were absent at this
order; subsequently it was explained~\cite{Gardi:2009qi,Becher:2009cu,Becher:2009qa} to be a direct consequence of the factorisation formula,
eq.~(\ref{softcolfac}), together with invariance under rescaling of
the four-velocities $\{\beta_i\}$. 
Invariance of the soft function alone
is broken for lightlike Wilson lines due to the appearance of
collinear singularities. It is restored however,
upon dividing by the eikonal jets, thus linking the breakdown of scale
invariance in the soft function to the jet function, and hence to the
cusp anomalous dimension (see ref.~\cite{Gardi:2009qi}). These
considerations lead to a differential equation for the soft anomalous
dimension, whose \emph{minimal solution} up to three-loop order is the
so-called {\it dipole
  formula}~\cite{Gardi:2009qi,Becher:2009cu,Becher:2009qa},
\begin{equation}
\Gamma_n^{\rm dip.}\left(\{p_i\},\{ {\bf T}_i\},\mu,\alpha_s\right)
=-\frac12\widehat{\gamma}_K(\alpha_s)\sum_{i<j}\log\left(\frac{-s_{ij}-i0}
{\mu^2}\right){\bf T}_i\cdot {\bf T}_j
+\sum_{i=1}^n\gamma_{J_i}(\alpha_s).
\label{gamdip}
\end{equation}
Here $\widehat{\gamma}_K$ is the cusp anomalous dimension~\cite{Korchemsky:1985xj,Korchemsky:1987wg,Moch:2004pa}, with the quadratic Casimir of the representation of the Wilson lines scaled out (see discussion below regarding the violation of Casimir scaling beyond three loops); 
$\gamma_{J_i}$ is an anomalous dimension associated with hard collinear
singularities, and is also known to three-loop order for both quark
and gluon jets~\cite{Moch:2005tm,Gehrmann:2010ue}. 

Refs.~\cite{Gardi:2009qi,Becher:2009cu,Becher:2009qa}  identified two classes of potential corrections to the soft anomalous dimension going beyond the dipole formula of eq.~(\ref{gamdip}). 
The first is related to the fact that
eq.~(\ref{gamdip}) contains the cusp anomalous dimension with all
colour dependence scaled out, thus assuming that {\it Casimir scaling}
holds to all orders. In fact, this breaks for the first time at
four-loop order due to the appearance of new colour structures, quartic Casimirs, as has very recently been shown explicitly in
refs.~\cite{Boels:2017skl,Boels:2017ftb,Moch:2017uml,Grozin:2017css}. This implies that the form of
eq.~(\ref{gamdip}) will have to be modified beyond three loops~\cite{Gardi:2009qi,Becher:2009cu,Becher:2009qa}. The
second source of corrections starts already at three loops, and
constitutes a homogeneous solution to the differential equation for
the soft anomalous dimension derived in ref.~\cite{Gardi:2009qi}. This
implies dependence on kinematics only through conformally invariant
cross ratios
\begin{equation}
\rho_{ijkl}\equiv\frac{(-s_{ij})(-s_{kl})}{(-s_{ik})(-s_{jl})}
=\frac{(\beta_i\cdot \beta_j)\,(\beta_k\cdot \beta_l)}
{(\beta_i\cdot \beta_k)\,(\beta_j\cdot\beta_l)},
\label{rhodef}
\end{equation}
such that the complete soft anomalous
dimension through three-loop order assumes the form
\begin{equation}
\Gamma_n(\{p_i\},\{ {\bf T}_i\},\mu,\alpha_s)=
\Gamma_n^{\rm dip.}(\{p_i\},\{ {\bf T}_i\},
\mu,\alpha_s)+\Delta_n(\{\rho_{ijkl}\},\{ {\bf T}_i\},\alpha_s)\,,
\label{gamnfull}
\end{equation}
where the correction 
\begin{equation}
\Delta_n(\{\rho_{ijkl}\},\{ {\bf T}_i\},\alpha_s) =\left(\frac{\alpha_s}{4\pi}\right)^3 \Delta_n^{(3)}(\{\rho_{ijkl}\},\{ {\bf T}_i\})+{\cal O}(\alpha_s^4)
\end{equation}
begins at three-loop order. Whether or not it is nonzero at this order
remained conjectural for a number of
years~\cite{Dixon:2009ur,Bret:2011xm,DelDuca:2011ae,Ahrens:2012qz,Naculich:2013xa,Caron-Huot:2013fea}.
Recently, however, it was calculated in
ref.~\cite{Almelid:2015jia}. Before we quote its form, let us note that for
any given four particles $\{i,j,k,l\}$, there are potentially 24 cross
ratios. However, eq.~(\ref{rhodef}) implies
\begin{equation}
\rho_{ijkl}=\rho_{jilk}=\rho_{klij}=\rho_{lkji},
\label{rhotrans1}
\end{equation}
which reduces the number of cross ratios to 6. In fact, further relations such as
\begin{equation}
\rho_{ijkl}=\frac{1}{\rho_{ikjl}},\quad
  \rho_{ijlk}\rho_{ilkj}=\rho_{ijkl}
\label{rhotrans2}
\end{equation}
can be used to write all the cross ratios in terms of just 2
independent cross ratios, which can be taken to be $\{\rho_{ijkl},
\rho_{ilkj}\}$. The explicit form of the three-loop
correction to the dipole formula can then be written
\begin{align}
\label{eq:expected_Delta}
& \Delta_n^{(3)}\left(\left\{\rho_{ijkl}\right\},\{{\bf T}_i\}\right) = 16\,f_{abe}f_{cde} 
\Big\{
-C\, \sum_{i=1}^n \ \sum_{\substack{{1\leq j<k\leq n}\\ j,k\neq i}}\left\{{\rm \bf T}_i^a,  {\rm \bf T}_i^d\right\}   {\rm \bf T}_j^b {\rm \bf T}_k^c  \\
&+\!\!\!{\sum_{1\leq i<j<k<l\leq n}}
\Big[
 {\rm \bf T}_i^a  {\rm \bf T}_j^b   {\rm \bf T}_k^c {\rm \bf T}_l^d   \, {\cal F}(\rho_{ikjl},\rho_{iljk}) 
+{\rm \bf T}_i^a  {\rm \bf T}_k^b {\rm \bf T}_j^c   {\rm \bf T}_l^d    
\, {\cal F}(\rho_{ijkl},\rho_{ilkj})
+ {\rm \bf T}_i^a   {\rm \bf T}_l^b  {\rm \bf T}_j^c    {\rm \bf T}_k^d 
\, {\cal F}(\rho_{ijlk},\rho_{iklj}) \Big]        \nonumber
\Big\}\,,
\end{align}
where 
\begin{equation}
C=\zeta_5+2\zeta_2\zeta_3,
\label{Cdef}
\end{equation}
and the explicit form of the function ${\cal F}$ can be most elegantly written by
introducing variables $\{z_{ijkl},\bar{z}_{ijkl}\}$ satisfying
\begin{equation}
z_{ijkl}\bar{z}_{ijkl}=\rho_{ijkl},\quad
(1-z_{ijkl})(1-\bar{z}_{ijkl})=\rho_{ilkj}.
\label{zdef} 
\end{equation}
With these definitions, one has
\begin{equation}
{\cal F}(\rho_{ijkl},\rho_{ilkj})=F(1-z_{ijkl})
-F(z_{ijkl}),
\label{calFdef}
\end{equation}
where
\begin{equation}
F(z)={\cal L}_{10101}(z)+2\zeta_2[{\cal L}_{001}(z)
+{\cal L}_{100}(z)].
\label{Fdef}
\end{equation}
The function ${\cal L}_w(z)$ (where $w$ is a word
composed of zeroes and ones) is a single-valued harmonic polylogarithm (SVHPL)~\cite{Brown:2004}.

\section{Single-valued harmonic polylogarithms and the Riemann sphere}
\label{sec:SVHPL}

The single-valued harmonic polylogarithms (SVHPLs)~\cite{Brown:2004,Remiddi:1999ew,Dixon:2012yy}  used to express the result in eq.~(\ref{Fdef}) are special combinations of harmonic polylogarithms~\cite{Remiddi:1999ew}
that are free of discontinuities, and thus single-valued, in the
kinematic region where $\bar{z}$ is equal to the complex conjugate of
$z$. This region is a subset of the so-called Euclidean region, where
all Mandelstam invariants are spacelike with $s_{ij}<0$. Unitarity of
massless scattering amplitudes dictates that they can only have
singularities due to the vanishing of Mandelstam invariants, which do
not occur in the Euclidean region of fixed angle scattering. Thus,
single-valuedness of the function $F$ reflects directly the analytic
structure of the underlying amplitude. 

It is remarkable that such a fundamental physics principle dictates the specific class of functions in terms of which the result for the soft anomalous dimension is expressed. This also provides the starting point for the bootstrap approach~\cite{Almelid:2017qju}, namely determining the result by 
constraining an ansatz, as an alternative to computing it directly. Ref.~\cite{Almelid:2017qju} started by explaining the geometrical origin of the kinematic dependence through the variables in eq.~(\ref{zdef}) and then how the class of weight-five SVHPLs of these variables can be identified as the relevant functions at three loops. Let us briefly review these considerations.

\begin{figure}[t]
\begin{center}
\includegraphics[scale=0.5]{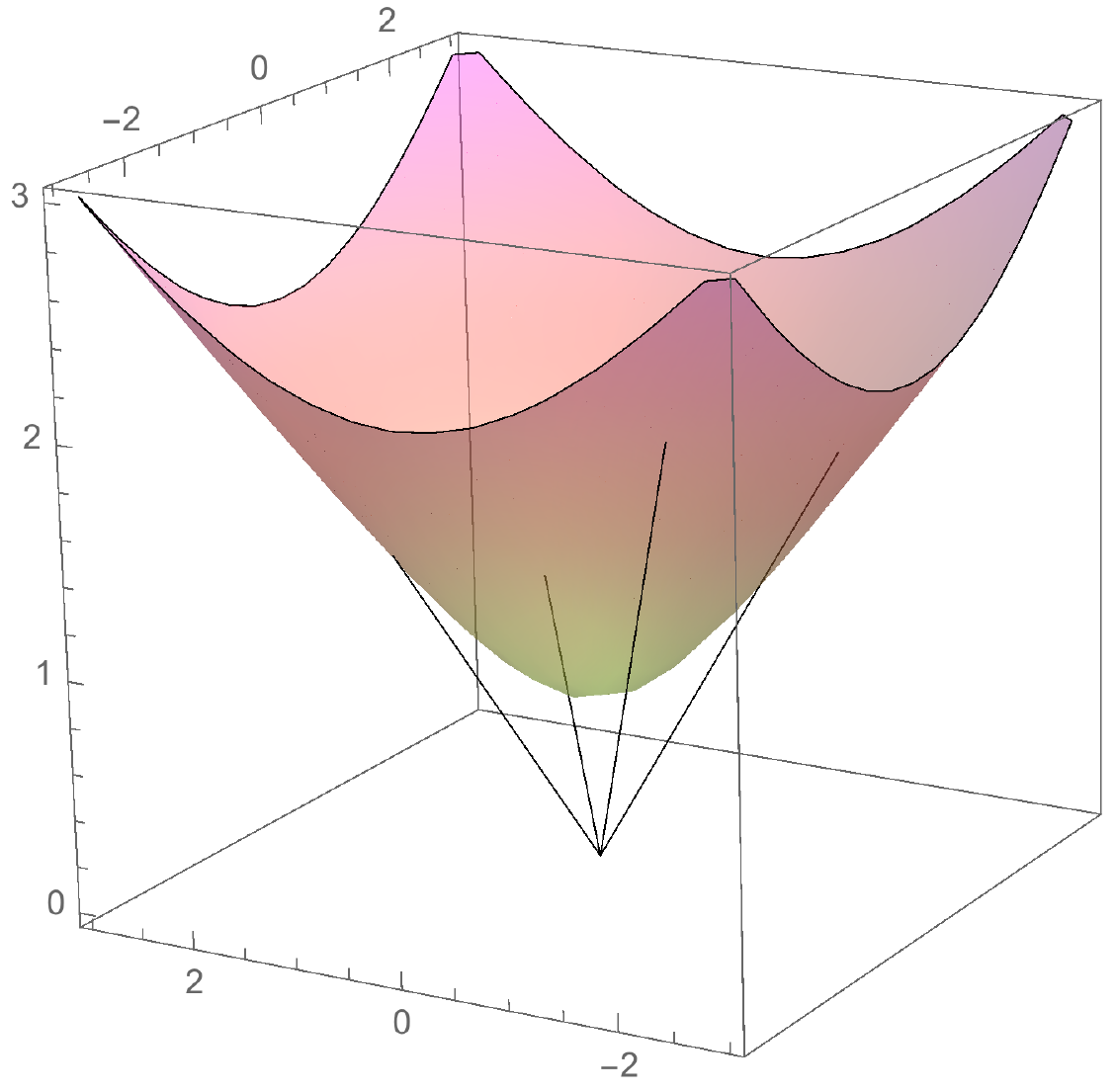}\qquad\qquad\qquad
\includegraphics[scale=0.4]{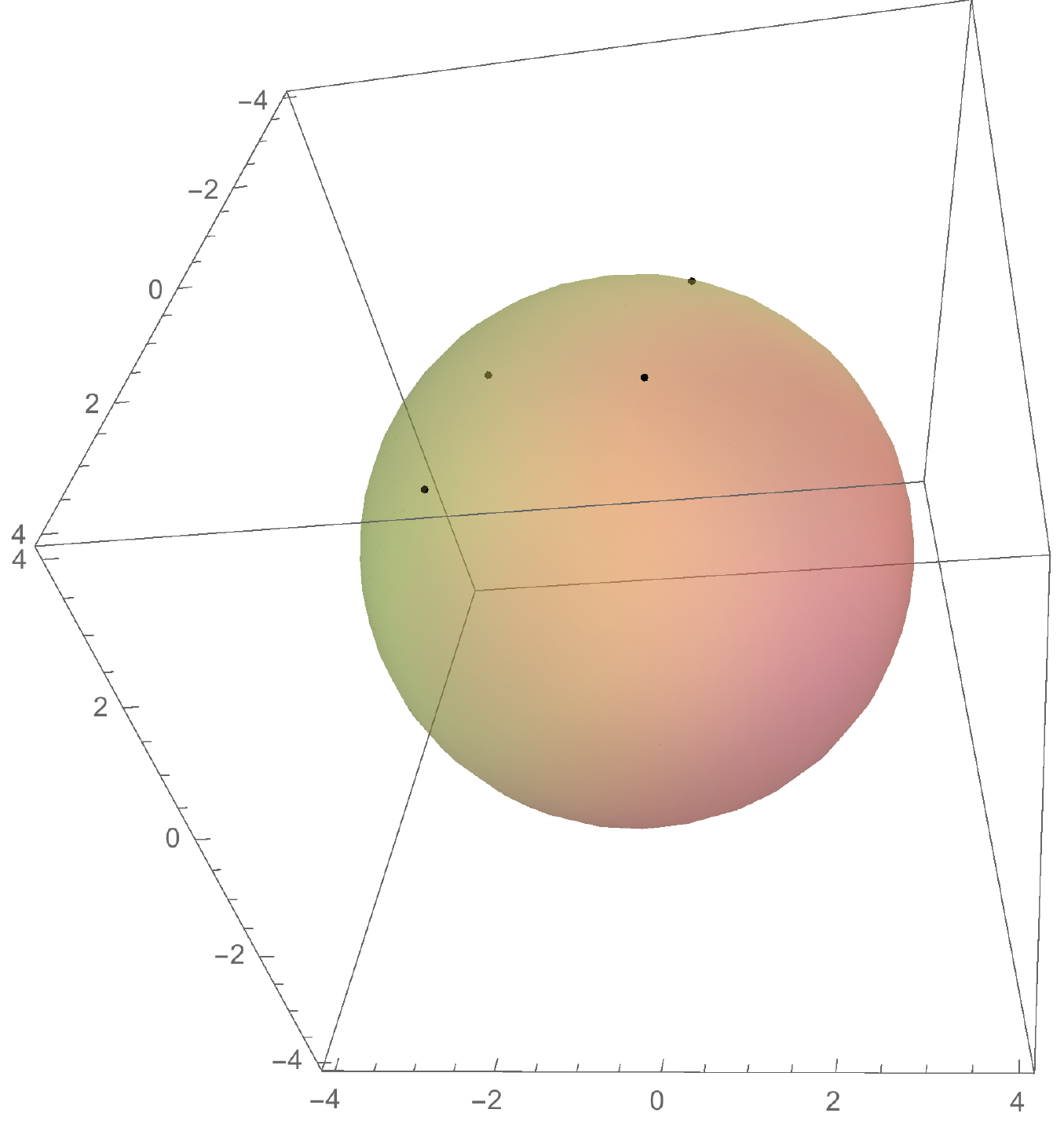}
\caption{\label{fig:Riemann_Sphere}Left -- the non-lightlike case: an illustration of the hyperbolic space (in one dimension less) with $\beta^0$ as the vertical axis, pierced by the $n$ velocity vectors; Right -- the lightlike case: the kinematics is described by $n$ points on a Riemann sphere, which is the boundary of the aforementioned hyperbolic space.}
\end{center}
\end{figure}

The correlator of Wilson lines in eq.~(\ref{Sdef}) defining the soft function is invariant under rescaling of individual Wilson-line velocity vectors, $\beta_i\to \kappa_i\beta_i$, and one can therefore fix the value of $\beta_i^2$ at will, without affecting the correlator. Upon considering all velocity vectors timelike and future pointing, and choosing a common value for $\beta_i^2$ for all $i$, one concludes that the kinematics is completely determined by a
set of $n$ points in three-dimensional hyperbolic space ${\mathbb H}^3$, namely the locus of all coordinates
\begin{equation}
\beta^\mu:\quad
 (\beta^0)^2-(\beta^1)^2-(\beta^2)^2 -(\beta^3)^2=R^2,\qquad \beta^0>0,
\end{equation}
for some constant $R$, as illustrated in fig.~\ref{fig:Riemann_Sphere} (left).
Feynman integrals are known to evaluate to iterated integrals, but the class of iterated integrals depending on $n$ points on ${\mathbb H}^3$ is yet to be fully understood. The situation is simpler however upon considering the lightlike limit relevant to massless scattering.
Parametrizing the velocity vectors by 
\begin{equation}
\label{upper_half}
\beta_i=
\left(1 + \frac{r_i^2 + z_i \bar{z}_i}{4} , \,\,
\frac{ z_i + \bar{z}_i}2 , \,\,
\frac{ z_i - \bar{z}_i}{2i},\,\, 
 1 - \frac{r_i^2 + z_i \bar{z}_i}{4} \right)
\end{equation}
where $\beta_i^2=r_i^2$ and  $z_i$ and $\bar{z}_i$ are complex conjugates,
the hyperbolic 3-space is represented by  the so-called upper half space model, ${\mathbb H}^3\simeq {\mathbb R}^2\times {\mathbb R}_+$. Now we can consider the lightlike limit, $r_i\to 0$, corresponding to the boundary of the aforementioned hyperbolic space. The velocity vector is then represented by a point on a Riemann sphere parametrized by the complex variable $z_i$. We thus see that while for massive scattering the kinematics is described by a configuration of $n$ points on ${\mathbb H}^3$, in massless scattering it is described by a configuration of $n$ points on a Riemann sphere, illustrated in fig.~\ref{fig:Riemann_Sphere} (right). The class of iterated integrals on the latter space,  ${\cal M}_{0,n}$, is well understood~\cite{Brown:2009qja}, as we now briefly discuss.

Let us first examine the relevant kinematic variables. In the coordinate system of (\ref{upper_half}) the product of two velocity vectors is given by
\begin{equation}
2\beta_i\cdot \beta_j=
|z_i-z_j|^2+r_i^2+r_j^2\,\,\,\xrightarrow{r_i,r_j\rightarrow \,0}\,\,\, |z_i-z_j|^2\,,
\label{betaibetaj}
\end{equation}
so in the lightlike limit, conformally invariant cross ratios (\ref{rhodef}) are given by
\begin{equation}
\rho_{ijkl}=\frac{(\beta_i\cdot \beta_j)(\beta_k\cdot \beta_l)}
{(\beta_i\cdot \beta_k)(\beta_j\cdot \beta_l)}\,\,\,
\xrightarrow{r_i,r_j,r_k,r_l\rightarrow \,0}\,\,\,
|{z}_{ijkl}|^2\,=\, {z}_{ijkl} {\bar{z}}_{ijkl}\,,\quad \text{with}\quad {z}_{ijkl}\equiv\frac{(z_i-z_j)\,(z_k-z_l) }{(z_i-z_k)\,(z_j-z_l)}.
\label{CICRz}
\end{equation}
Furthermore, we can use the symmetries of the sphere, namely an ${\rm
  SL}(2,{\mathbb C})$ invariance, to
fix the positions of three of the $n$ points. Specifically, considering a subset of
four Wilson lines $\{i,j,k,l\}$,  one may choose
\begin{equation}
z_i=z_{ijkl},\quad z_j=0,\quad z_k=\infty,\quad z_l=1,
\label{zchoice}
\end{equation}
so that the only nontrivial position $z_i$ on the Riemann sphere may
be identified with the variable of eqs.~(\ref{zdef}). Specifically, three-loops webs span up to four Wilson lines (see~\cite{Gardi:2013ita,Almelid:2015jia,Gardi:2016ttq}), so it is clear at the outset that the function ${\cal F}$ can only depend on \emph{one} complex variable, $z_{ijkl}$, independently of the number of scattered particles $n$. 

Having identified the kinematic variables, we may use the fact that the nature of iterated integrals on the space of configurations of $n$ points on the sphere, ${\cal M}_{0,n}$ is completely known: they can always be expressed as linear combinations of products of multiple polylogarithms, and the coefficients of this linear combination are rational functions~\cite{Brown:2009qja}. 
At this point we can use further considerations to restrict the class of functions (see ref.~\cite{Almelid:2017qju} for further details):
\begin{itemize} 
\item{} The physical restriction of having strictly no branch points in the Euclidean region, where all $s_{ij}<0$, selects the class of \emph{single valued} polylogarithms. 
\item{} Given that three-loops webs span up to four Wilson lines the relevant functions depend on 
a single (complex) variable $z_{ijkl}$. This uniquely identifies them as single-valued \emph{harmonic} polylogarithms. This is because singularities appear only when two of the points coincide, e.g. $z_i\to z_j$, corresponding to collinear limits; this dictates the symbol alphabet:
$\left\{z_{ijkl},\bar{z}_{ijkl}, 1-z_{ijkl}, 1-\bar{z}_{ijkl}\right\}$.
\item{} While a priori both the polylogarithms and their rational coefficients depend on the kinematic variables $z_{ijkl}$, the fact the three-loop $\Delta_n^{(3)}$ is the same in all gauge theories, along with the conjectural property that ${\cal N}=4$ amplitudes (in the maximally-helicity-violating sector) are pure functions of uniform (maximal) weight, leads to the expectation that ${\cal F}$ should be pure and of uniform weight $5$.
\end{itemize}
Having identified the relevant functions it is possible to set up a bootstrap procedure, as was done in ref.~\cite{Almelid:2017qju}. We will describe this after reviewing the available information from the high-energy and collinear limits.

\section{Soft singularities in the high-energy limit}

\begin{figure}[t]
\begin{center}
\includegraphics[scale=0.8]{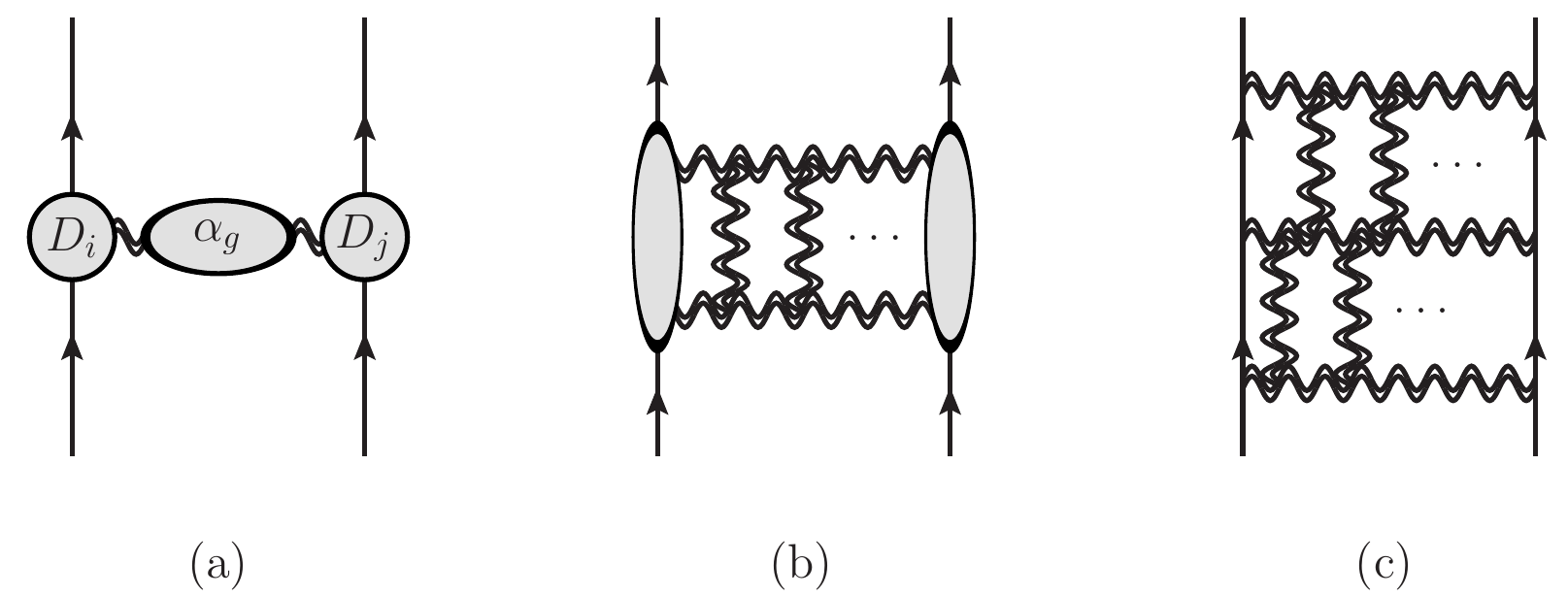}
\caption{\label{Regge_exchanges}From left to right, exchange of one, two and three Reggeized gluons, respectively. We draw the Reggeized gluons as double wavy lines. Single Reggeon exchange in the first diagram contributes at LL accuracy to the real part of the amplitude, while two-Reggeon exchange in the second diagram contributes at NLL accuracy to the imaginary part. Last, three Reggeons exchange starts contributing to the real part at NNLL accuracy. The shaded blobs in the first and second diagram account for single- and two-Reggeon impact factors, which give additional contributions at subleading logarithmic accuracy to these diagrams.}
\end{center}
\end{figure}

One of the most important and theoretically interesting limits to consider is the high-energy limit  of parton scattering where the centre-of-mass energy $s$ is much larger than the momentum transfer $-t$, see refs.~\cite{Sotiropoulos:1993rd,Korchemsky:1993hr,Korchemskaya:1996je,Korchemskaya:1994qp,DelDuca:2001gu,DelDuca:2013ara,DelDuca:2014cya,Bret:2011xm,DelDuca:2011ae,Caron-Huot:2013fea,Caron-Huot:2017fxr,Caron-Huot:2017zfo}. 
Specifically, let us focus here on nearly forward  $2 \to 2$ parton scattering, which can be scattering of two gluons, two (anti-) quarks, or a gluon and an (anti-) quark. At tree-level this limit is dominated by the $t$-channel gluon-exchange diagram. At higher orders in perturbation theory it is corrected by large logarithms that, to leading logarithmic accuracy, simply exponentiate by modifying the $t$-channel gluon propagator according to
\begin{equation}
\label{LL_Regge}
\frac{1}{t}\to \frac{1}{t}\, \left(\frac{s}{-t}\right)^{\alpha(t)}\,,
\end{equation}
dubbed ``Gluon Reggeization'', where $\alpha(t)$ is the gluon Regge trajectory, which is infrared-singular. 
Gluon Reggeization can be extended to next-to-leading logarithms (NLL) for the real part of the amplitude by including ${\cal O}(\alpha_s^2)$ corrections in the trajectory and in addition
introducing impact factors that describe the interaction of the Reggeized gluon
with the target and the projectile. These impact factors depend on the scattered particle and on the momentum transfer $-t$, but are independent of the energy. This factorization is represented by diagram (a) in fig.~\ref{Regge_exchanges}, and is referred to as Regge-pole high-energy factorization.    

The infrared singularity structure implied by gluon Reggeization according to eqs.~(\ref{LL_Regge}) and its extension to NLL for the real part of the amplitude,  is fully consistent with soft-gluon exponentiation where the soft anomalous dimension is taken to be the dipole formula of eq.~(\ref{gamdip}). 
However, this simple picture does not extend beyond NLL due to \emph{multiple 
Reggeized gluon exchange}, which forms Regge cuts.

The possibility to compute directly logarithmically-enhanced terms associated with multiple Reggeized gluons exchange in partonic scattering is an important theoretical development of the recent years, starting with the seminal paper of Simon Caron-Huot~\cite{Caron-Huot:2013fea}. In this framework Reggeized gluons are sourced by the logarithm of an infinite Wilson-line operator along the trajectory of the target or projectile, defined at a given point in the  $(2-2\epsilon)$ dimensional transverse space. These effective degrees of freedom are governed by rapidity evolution equations, namely BFKL and its non-linear generalizations by Balitsky-JIMWLK~\cite{Kuraev:1977fs,Balitsky:1978ic,Lipatov:1985uk,Mueller:1993rr,Mueller:1994jq,Balitsky:1995ub,Balitsky:1998kc,Kovchegov:1999yj,JalilianMarian:1996xn,JalilianMarian:1997gr,Iancu:2001ad}, which can be solved order-by-order in perturbation theory~\cite{Caron-Huot:2017fxr,Caron-Huot:2017zfo}.
 
\begin{figure}[t]
\begin{center}
\includegraphics[scale=0.7]{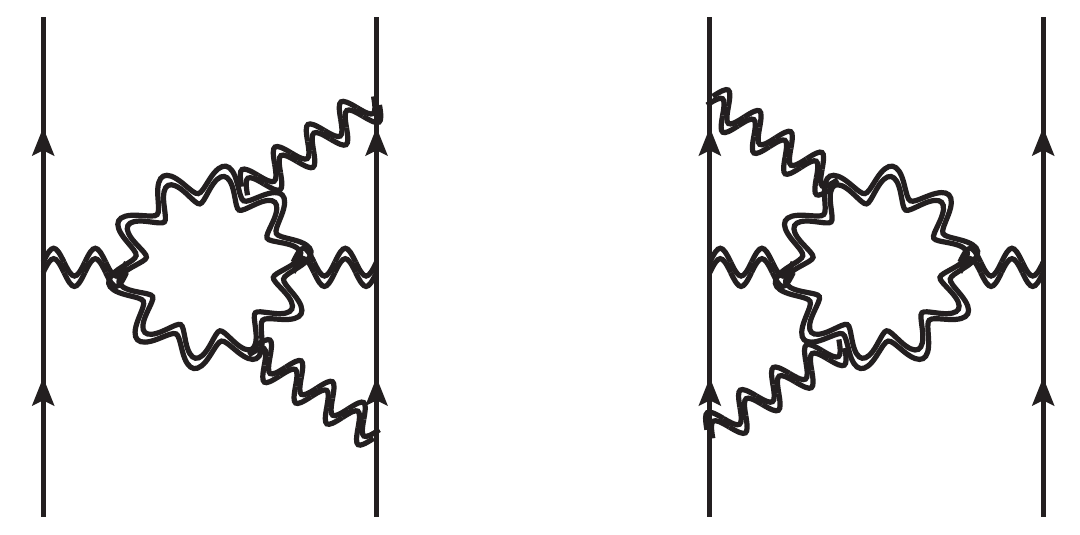}\qquad\quad
\includegraphics[scale=0.7]{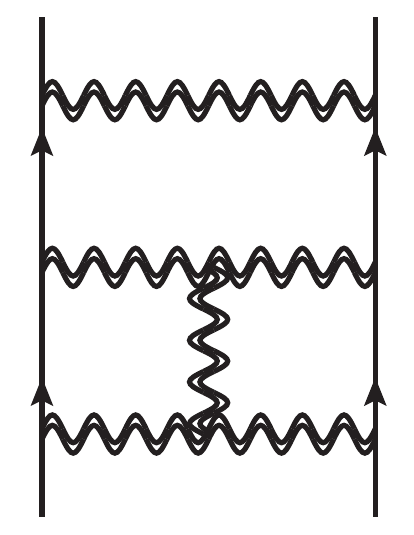}
\caption{\label{Odd_three_loop_Regge_exchanges}Three loop contributions associated with $1\to 3$, $3\to 1$ and $3\to 3$ Reggeized gluon exchanges. These contributions were evaluated in Ref.~\cite{Caron-Huot:2017fxr}.}
\end{center}
\end{figure}

Considering the real part of the amplitude at NNLL accuracy, Regge-cut contributions were computed explicitly in ref.~\cite{Caron-Huot:2017fxr} through three loops (see also~\cite{Fadin:2016wso,Fadin:2017nka}). At this order one encounters a couple of new effects including the evolution of a state consisting of three Reggeized gluons and the mixing between the latter and 
a single Reggeized gluon, as shown in fig.~\ref{Odd_three_loop_Regge_exchanges}.
Once this mixing is taken into account one can also define the three-loop gluon Regge trajectory~\cite{Caron-Huot:2017fxr} (the latter is known in ${\cal N}=4$ super Yang-Mills thanks to the computation of the amplitude in ref.~\cite{Henn:2016jdu}, but not in QCD). 

The imaginary part of the amplitude is affected already at NLL accuracy by Regge cuts due to the exchange of two Reggeized gluons~\cite{Caron-Huot:2013fea} (see also~\cite{DelDuca:2001gu,Bret:2011xm,DelDuca:2011ae,DelDuca:2013ara,DelDuca:2013dsa,DelDuca:2014cya}) which form the famous BFKL ladder graphs shown in diagram (b) in fig.~\ref{Regge_exchanges}. The corresponding infrared singularities were computed very recently to all orders in perturbation theory~\cite{Caron-Huot:2017zfo}.  

Let us now summarise the state-of-the-art knowledge of infrared singularities in the high-energy limit of 
$2\to 2$ QCD scattering according to refs.~\cite{Caron-Huot:2017fxr,Caron-Huot:2017zfo}. To this end we write the soft anomalous dimension in this limit by collecting towers of increasing logarithmic accuracy as follows:
\begin{equation}
{\Gamma} \left(\alpha_s\right) = 
{\Gamma}_{\rm LL}\left(\alpha_s,L\right) 
+ {\Gamma}_{\rm NLL}\left(\alpha_s,L\right) 
+ {\Gamma}_{\rm NNLL} \left(\alpha_s,L\right) 
+ \ldots\,
\end{equation}
where  ${\Gamma}_{{\rm N}^k{\rm LL}}$ collects all terms with $\alpha_s^n L^{n-k}$. We choose the high-energy logarithm, $L$, to be of even signature (i.e. symmetric under $u$ to $s$ exchange):
\begin{equation}
L=\frac12\left(\ln \frac{-s-i0}{-t}+\ln \frac{-u-i0}{-t}\right)=\ln \left|\frac{s}{t}\right| -i\frac{\pi}{2}\,,
\end{equation}
so that terms with odd (even) signature in the amplitude are real (imaginary) (see section 2 in ref.~\cite{Caron-Huot:2017fxr}). Given that the tree-level hard amplitude is itself odd, real (imaginary) terms in the soft anomalous dimension are even (odd).
The leading-logarithmic anomalous dimension is one-loop exact,
\begin{equation}
{\Gamma}_{\rm LL}\left(\alpha_s,L\right) =\frac{\alpha_s}{\pi}L {\bf T}_t^2\,,
\end{equation}
and the real part of the NLL one is two-loop exact:
\begin{equation}
{\Gamma}_{\rm NLL}^{(+)} =  \frac{\alpha_s}{\pi} \sum_{i=1}^{2} \left( \frac{\widehat{\gamma}^{(1)}_{K}}{2} C_i \log \frac{-t}{\lambda^2} 
+ 2{\gamma}^{(1)}_{J_i} \right)
\,\,+\,\,
\left(\frac{\alpha_s}{\pi}\right)^2 \frac{\widehat{\gamma}^{(2)}_{K}}{2} L \, \Tt\,,
\end{equation}
as can be seen from eq.~(\ref{gamdip}) (the coefficients of the cusp and collinear anomalous dimensions are summarised in Appendix A of ref.~\cite{Caron-Huot:2017fxr}).
The full soft anomalous dimension at NLL in the high-energy limit, ${\Gamma}_{\rm NLL}$, also has a signature-odd imaginary part, ${\Gamma}_{\rm NLL}^{(-)}$.
At one loop it can be deduced from eq.~(\ref{gamdip}) (see refs.~\cite{Bret:2011xm,DelDuca:2011ae,DelDuca:2013ara,DelDuca:2013dsa,DelDuca:2014cya}), 
\begin{equation}
{\Gamma}_{\rm NLL}^{(-)} = i\pi \frac{\alpha_s}{\pi} \,\Tsu\,+
{\cal O} (\alpha_s^4)\,,
\end{equation}
but it also receives corrections starting at four loops~\cite{Caron-Huot:2013fea}. These have recently been computed to all orders in ref.~\cite{Caron-Huot:2017zfo}:
\begin{equation}
{\Gamma}_{\rm NLL}^{(-)} = i\pi \frac{\alpha_s}{\pi} \,G\left(\frac{\alpha_s}{\pi}L\right)\Tsu\,,
\end{equation}
where 
\begin{equation}
G(x) \equiv \sum_{l=1}^{\infty} G^{(l)} x^{l-1}\quad\text{with}\quad
G^{(l)} \equiv 
\frac{1}{(l-1)!}\left[ \frac{(C_A-\Tt)}{2}\right]^{l-1} \left.\left( 1 - \frac{C_A}{C_A -\Tt}R(\epsilon)  \right)^{-1}\right\vert_{\epsilon^{l-1}} 
\label{G}
\end{equation}
and where
\begin{equation}
R(\epsilon) = 
\frac{\Gamma^{3}(1-\epsilon)\Gamma(1+\epsilon)}{\Gamma(1-2\epsilon)} -1  
= -2\zeta_3 \, \epsilon^3 -3\zeta_4 \, \epsilon^4 -6\zeta_5 \epsilon^5
-\left(10 \zeta_6-2\zeta^2_3 \right) \epsilon^6 + {\cal O}(\epsilon^7)\,.
\end{equation}
We can see that new colour structures emerge at higher orders, depending on ${\bf T}_t^2$ and $C_A$. It is not surprising that the result has maximal weight at any order, as it applies to
${\cal N}=4$ super Yang-Mills. 
What is interesting, and unusual, is that this anomalous dimension sums up to an entire function of the coupling times the logarithm, so it has an infinite radius of convergence -- this is due to the factorial suppression of the coefficients in eq.~(\ref{G}). We refer the interested reader to ref.~\cite{Caron-Huot:2017zfo} for further details.
 
At the next logarithmic order, NNLL, three-loop results are known explicitly~\cite{Caron-Huot:2017fxr}:
\begin{equation}
\label{NNLL_results}
{\Gamma}_{\rm NNLL}^{(+)} ={\cal O} (\alpha_s^4)\,;\qquad 
{\Gamma}_{\rm NNLL}^{(-)} = i\pi\left[ \frac{\zeta_3}{4}(C_A-\Tt)^2 \,\left(\frac{\alpha_s}{\pi}\right)^3 L+
{\cal O} (\alpha_s^4)
\right]\Tsu\,,
\end{equation}
but higher orders remain so far unknown. The fact that the real part, ${\Gamma}_{\rm NNLL}^{(+)}$, vanishes at three loops was deduced from the computation~\cite{Caron-Huot:2017fxr} based on Balitsky-JIMWLK equation. In other words, at this logarithmic order the rapidity evolution equation reproduces precisely the poles that are generated by the dipole formula of eq.~(\ref{gamdip}). 
This was shown to be consistent with the result for the three-loop soft anomalous dimension~\cite{Almelid:2015jia}, which does not produce any new contributions proportional to $L^1$ upon taking the high-energy limit. 
The non-vanishing imaginary value ${\Gamma}_{\rm NNLL}^{(-)}$ quoted in eq.~(\ref{NNLL_results}) was deduced by taking the high-energy limit of the soft anomalous dimension of~ref.~\cite{Almelid:2015jia}; a corresponding direct computation in the high-energy limit at this order would require accounting for  the exchange of four Reggeised gluons.


To summarise: we have seen that significant progress was achieved recently in our ability to compute logarithmic corrections in the high-energy limit of $2\to 2$ scattering associated with the exchange of multiple Reggeized gluons by iteratively solving the BFKL and JIMWLK rapidity evolution equations. 
The interplay between the exponentiation of soft singularities owing to infrared factorization and the exponentiation of high-energy logarithms by means of rapidity evolution equations has proven to be highly insightful in both directions. 
Specifically, at three loops, the absence of logarithmically enhanced contributions with $\alpha_s^3 L^k$ for any $k\geq 1$  for the real part, and any $k\geq 2$ in the imaginary part, provides a very strong constraint on the structure of the soft anomalous dimension.

\section{Collinear limits}

Next, let us consider the collinear limit of an $n$-parton amplitude following refs.~\cite{Becher:2009qa,Dixon:2009ur,Almelid:2017qju}. 
This is the limit where two of the external hard partons become collinear. In general, this limit gives rise to another factorization property we have not discussed so far: this is the so-called collinear or splitting-amplitude factorization. 

Working in the kinematic region where all $n$ coloured particles carrying momenta $\{p_k\}$ are outgoing, we consider the limit in which two of these partons, $i$ and $j$, become collinear.  In this limit $s_{ij}=2p_i\cdot p_j\rightarrow
0$, resulting in kinematic divergences proportional to $1/s_{ij}$. It is well known that for final-state collinear partons\footnote{In
  the case of a space-like splitting, factorisation is
  violated~\cite{Catani:2011st,Forshaw:2012bi}.} these divergences factorise~\cite{Bern:1999ry,Kosower:1999xi,Feige:2014wja,Catani:2003vu}, such that one may write
\begin{equation}
{\cal A}_n(p_1,p_2,\{p_j\};\mu,\epsilon,\alpha_s)
\xrightarrow{1\parallel 2}{\bf Sp}(p_1,p_2,{\bf T}_1,{\bf T}_2;\mu,\epsilon,\alpha_s)\,
{\cal A}_{n-1}(P,\{p_j\},\mu,\epsilon,\alpha_s).
\label{colfac}
\end{equation}
Without loss of generality, we have taken particles 1 and 2 collinear,
where $\{p_j\}$, $j=3\ldots n$ denotes the set of remaining
momenta. The right-hand side contains the $(n-1)$-particle amplitude in
which the momenta $p_1$ and $p_2$ are replaced by the sum $P=p_1+p_2$,
multiplied by a universal {\it splitting function} ${\bf Sp}$, which collects
all singular contributions to the amplitude due to particles
1 and 2 becoming collinear. 
\begin{figure}[htb]
\begin{center}
\includegraphics[scale=0.7]{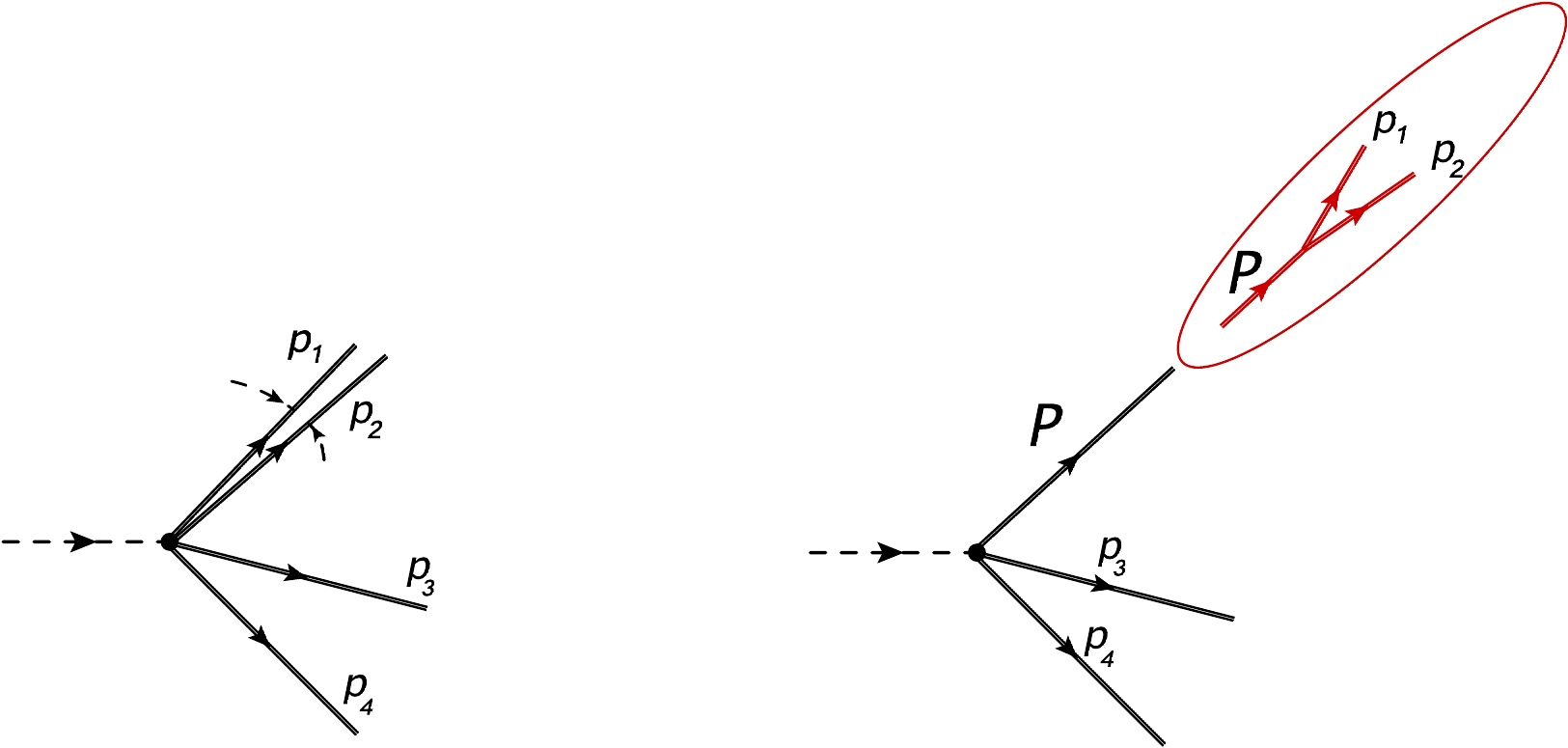}
\caption{\label{Collinear_limit}Left: An $n$ parton amplitude (in this case $n=4$) with $p_1$ nearly collinear to $p_2$; Right: The same, factorized into a corresponding $(n-1)$ parton amplitude where partons $1$ and $2$ have been replaced by a single particle carrying momentum $P=p_1+p_2$ and colour charge ${\bf T}={\bf T}_1+{\bf T}_2$, times a splitting amplitude describing the collinear pair.}
\end{center}
\end{figure}
A central property of the splitting amplitude ${\bf Sp}$ is that it depends only on the degrees of freedom (momenta, colour and helicities) of the collinear particles, and is entirely independent of the rest of the process. 

\begin{figure}[t]
\begin{center}
\includegraphics[scale=0.7]{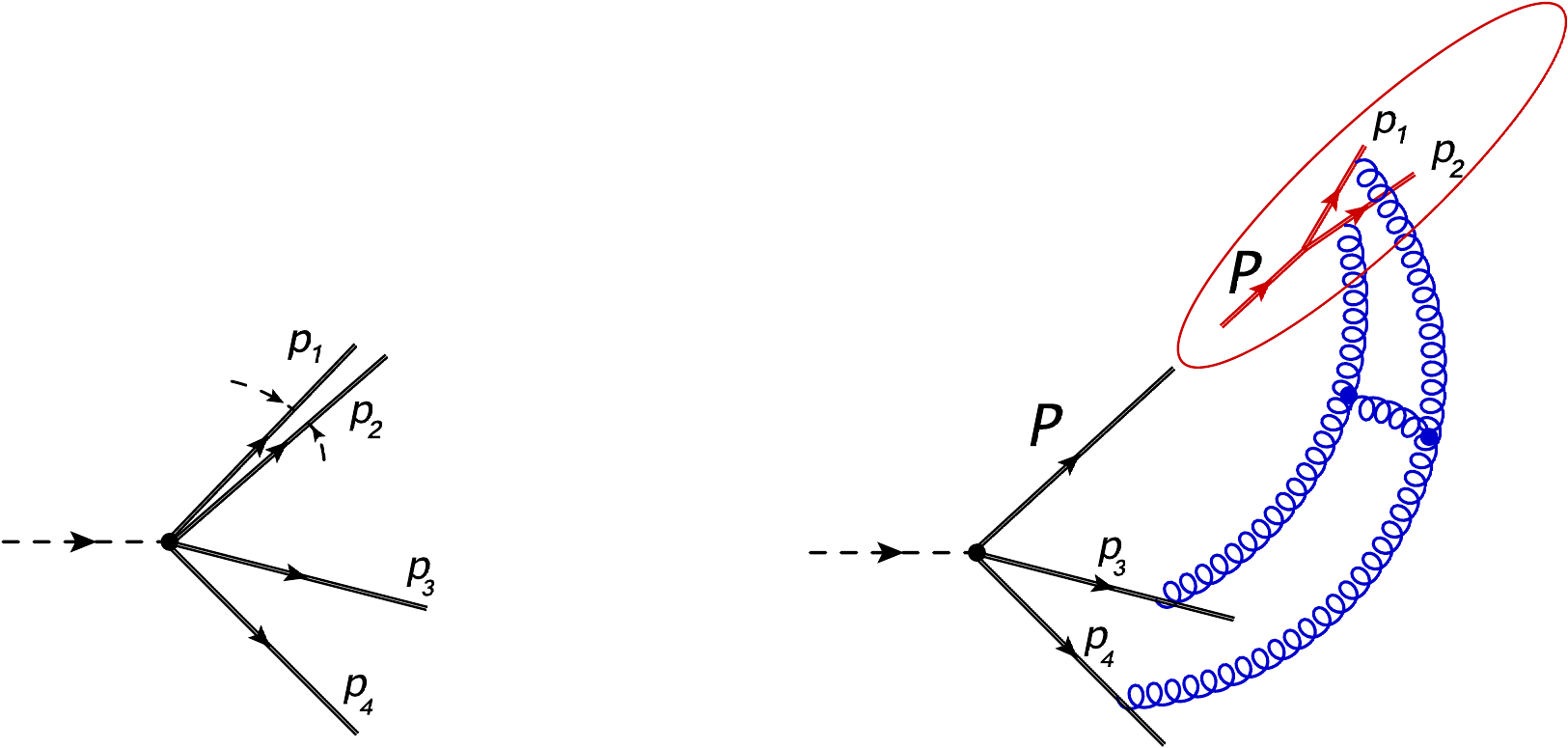}
\caption{\label{Collinear_limit_gluons}As in fig.~\ref{Collinear_limit}, but where the the collinear pair interacts with the rest of the process via a three-loop web, that on the face of it, could violate collinear factorization. In effect, such a violation does not occur (when all partons are ourgoing) owing to intricate cancellations involving colour conservation.}
\end{center}
\end{figure}

It can be shown that the singularities of the splitting amplitude are controlled by an anomalous dimension which is the difference between the soft anomalous dimensions of the $n$ parton amplitude and the $(n-1)$ one~\cite{Becher:2009qa,Dixon:2009ur,Almelid:2017qju}:
\begin{equation}
\begin{split}
\Gamma_{\bf Sp}&(p_1,p_2,{\bf T}_1,{\bf T}_2,\mu_f,\alpha_s(\mu_f^2))\\
&\,\equiv \Gamma_n(p_1,p_2,\{p_j\},{\bf T}_1,{\bf T}_2,\{{\bf T}_j \},\mu_f,\alpha_s(\mu_f^2))
-\Gamma_{n-1}(P,\{p_j\},{\bf T},\{ {\bf T}_j \},\mu_f,\alpha_s(\mu_f^2))\,.
\label{GamSpdef}
\end{split}
\end{equation}
Given that the quantity on the left-hand side of this equation depends
manifestly on the quantum numbers of partons 1 and 2 only, this must
also be true for the right-hand side. 
At the level of the dipole formula $\Gamma_{\bf Sp}$ can be computed and upon using colour conservation it immediately satisfies the requirement that it only involves the degrees of freedom of the collinear pair.
At three loops there is potential for this to break down, since there are diagrams that explicitly correlate the collinear pair and the rest of the process, as illustrated in fig.~\ref{Collinear_limit_gluons}. 
This leads to an important constraint on the $\Delta_n$ in eq.~(\ref{gamnfull}), namely that the difference
\begin{equation}
\Delta^{(3)}_{\bf Sp}({\bf T}_1,{\bf T}_2)=\left[
\Delta^{(3)}_n(\{\rho_{ijkl}\},{\bf T}_1,{\bf T}_2,\{{\bf T}_j\})
-\Delta^{(3)}_{n-1}(\{\rho_{ijkl}\},{\bf T},\{{\bf T}_j\})
\right]_{p_1\parallel p_2}
\label{delcoll}
\end{equation}
can only depend on the quantum numbers of the two particles
that are becoming collinear. Note that the right-hand side of eq.~\eqref{delcoll} is
evaluated in the limit where $p_1$ and $p_2$ have become collinear. 
Since $\Delta^{(3)}_{\bf Sp}({\bf T}_1,{\bf T}_2)$ is expected to be universal, it can be extracted from any $n$-point amplitude. Special simplification occurs for $n=3$, since in this case $\Delta^{(3)}_{(n-1)}$ vanishes and it follows that 
\begin{equation}\begin{split}
\label{DelSp3}
\Delta^{(3)}_{\bf Sp}({\bf T}_1,{\bf T}_2)&\,=\left[\Delta^{(3)}_3(-{\bf T}_1-{\bf T}_2,{\bf T}_1,{\bf T}_2)-\Delta_2^{(3)}({\bf T}_1,{\bf T}_2)
\right]_{p_1\parallel p_2}\\
&\,=\left.\Delta^{(3)}_3(-{\bf T}_1-{\bf T}_2,{\bf T}_1,{\bf T}_2)\right|_{p_1\parallel p_2}\,,
\end{split}
\end{equation}
but since there are no conformally invariant cross ratios that one can form
from three particle momenta, the right-hand side evaluates to a
constant.

Indeed, using the result of eqs.~(\ref{eq:expected_Delta}) through (\ref{Fdef}) we obtain:
\begin{equation}
\label{Split_delta}
\Delta_{\rm \bf Sp}^{(3)}=\left.(\Delta_n^{(3)}-\Delta_{n-1}^{(3)})\right\vert_{1\parallel 2}=-24 \left(\frac{\alpha_s}{4\pi}\right)^3\, (\zeta_5+2\zeta_2\zeta_3)\,\left[f^{abe}f^{cde}\left\{{\rm \bf T}_1^a,  {\rm \bf T}_1^c\right\}\left\{{\rm \bf T}_2^b,  {\rm \bf T}_2^d\right\} +\frac12C_A^2  {\rm \bf T}_1\cdot {\rm \bf T}_2\right]\,.
\end{equation}
Note that $\Delta_n^{(3)}$ and $\Delta_{n-1}^{(3)}$ are separately dependent on both colour and kinematics of the rest of the process. Their difference reduces to (\ref{Split_delta}) through intricate cancellations involving colour conservation and an interplay between the constant $C$ and kinematic-dependent function ${\cal F}$. 
This can be regarded as a highly non-trivial confirmation of the properties and universality of splitting amplitudes, as well as a check of the calculation itself. Conversely, the requirement that the right hand side of eq.~(\ref{delcoll}) only depends on the degrees of freedom of the collinear pair can be used as a constraint on $\Delta_n^{(3)}$ as was done in ref.~\cite{Almelid:2017qju}, feeding into the determination of this quantity by bootstrap.

\section{Bootstrapping the three-loop soft anomalous dimension}

The bootstrap procedure of ref.~\cite{Almelid:2017qju} to determine the three-loop soft anomalous dimension starts off with two ingredients. The first is the  colour structure, which is known to be the set of fully-connected colour factors based on the non-Abelian exponentiation theorem for multiple Wilson lines~\cite{Gardi:2013ita}. The second is the identification of the relevant class of functions as weight-five SVMPLs as established in~ref.~\cite{Almelid:2017qju} and briefly reviewed in section~\ref{sec:SVHPL} above.

Considering the full set of connected colour factors, 
the most general form that $\Delta_n^{(3)}$ can take is:
\begin{align}
\begin{split}
\label{eq:Delta_general_color}
\Delta_n^{(3)}&= \sum_{\{ i,j,k,l \}} f_{abe}f_{cde}\, {\bf T}_i^a {\bf T}_j^b {\bf T}_k^c {\bf T}_l^d \,A_{ijkl}
+\sum_{\{ i,j,k \}}  f_{abe} f_{cde} \left\{{\bf T}_i^a,{\bf T}_i^d\right\} {\bf T}_j^b {\bf T}_k^c \,B_{ijk}\\
&+\sum_{\{ i,j,k \}}  f_{abc}{\bf T}_i^a{\bf T}_j^b{\bf T}_k^c \,C_{ijk}
+\sum_{\{ i,j \}}   f_{abe} f_{cde} \left\{{\bf T}_i^a,{\bf T}_i^d\right\} \left\{{\bf T}_j^b,{\bf T}_j^c\right\} \,D_{ij}
+\sum_{\{ i,j \}}  {\bf T}_i\cdot {\bf T}_j \,E_{ij}\ ,
\end{split}
\end{align}
where the sums run over all sets $\{ i,j,\ldots \}$ of distinct Wilson
lines, and the coefficients $A_{ijkl}$, $B_{ijk}$ etc. are functions
of the four-velocities of the Wilson lines entering each colour factor.
Equation~\eqref{eq:Delta_general_color}, however, is still largely
over-complete. First, it was shown in
ref.~\cite{Gardi:2009qi,Becher:2009cu,Becher:2009qa} that colour
tripoles of the form $f_{abc}{\bf T}_i^a{\bf T}_j^b{\bf T}_k^c$ are
absent at any loop order (a corresponding kinematic function would violate rescaling invariance) and so we must have $C_{ijk}=0$ in
eq.~\eqref{eq:Delta_general_color}. Second, the soft anomalous dimension, and hence 
$\Delta_n^{(3)}$, depends on the colour generators subject to colour conservation, namely
\begin{equation}
\label{eq:colour_conservation} 
\left(\sum_{i=1}^n{\bf
  T}_i^a\right){\cal H}_n = 0\,.  
\end{equation}
In refs.~\cite{AlmelidPhD,Gardi:2016ttq,longinprep} the role of colour
conservation in the context of $\Delta_n^{(3)}$ was analysed in
detail, and a basis of colour structures that are independent after
imposing eq.~\eqref{eq:colour_conservation} was worked out. In this basis the general ansatz takes the much simpler form:
\begin{align}
\begin{split}
\label{eq:ansatz_colour}
\Delta_n^{(3)}&= \sum_{\{i,j,k,l\}} f_{abe}f_{cde}\, {\bf T}_i^a {\bf T}_j^b {\bf T}_k^c {\bf T}_l^d \,A_{ijkl}(\rho_{ikjl},\rho_{iljk})
\,-16\,C\,\sum_{i=1}^n \ \sum_{\substack{{1\leq j<k\leq n}\\ j,k\neq i}}  f_{abe} f_{cde} \left\{{\bf T}_i^a,{\bf T}_i^d\right\} {\bf T}_j^b {\bf T}_k^c\,\,.
\end{split}
\end{align}
Finally, using Bose symmetry one can show that eq.~(\ref{eq:ansatz_colour}) takes the form of  eq.~(\ref{eq:expected_Delta}) with ${\cal F}$ given by eq.~(\ref{calFdef}), where we now regard $F(z)$ as an unknown function and $C$ as an unknown constant. 

The next step is to write a general ansatz for these in terms of weight-five SVHPLs.  
A general ansatz for $F(z)$ consists of 32 weight-five SVHPLs, 8
weight-three SVHPLs multiplied by $\zeta_2$, 4 weight-two SVHPLs
multiplied by $\zeta_3$, 2 weight-one SVHPLs multiplied by $\zeta_4$,
and finally a general linear combination of the 2 weight-five
constants $\zeta_5$ and $\zeta_{2}\zeta_3$, all with rational
prefactors. This gives 48 distinct terms. Similarly, a general ansatz
for the constant $C$ consists of a general linear combination of the
2 independent weight five multiple zeta values, with rational prefactors. However, this
na\"ive ansatz is overly large for two reasons. First, because of the
invariance under the interchange $z_{ijkl}\leftrightarrow\bar{z}_{ijkl}$ and the way
this symmetry acts on SVHPLs via reversal of words ${\cal L}_w(\bar{z})={\cal L}_{\tilde{w}}(z)+\ldots$, only palindromic combinations of weight five
SVHPLs need to be considered.
Second, the function $F(z_{ijkl})$ only appears in our ansatz, eq.~(\ref{eq:expected_Delta}), via
the differences:
\begin{align}
F(1-1/z)-F(1/z),\quad F(1-z)-F(z),\quad F(1/(1-z))-F(z/(z-1)).
\label{Fdiffs}
\end{align}
Taking these considerations into account results in an ansatz for $F(z)$ involving only 11 parameters:
\begin{align}\label{ansatz1}
F(z)&=a_1{\cal L}_{00000}+a_2{\cal L}_{00100}+a_3{\cal L}_{10001}
+a_4{\cal L}_{10101}+a_5\left({\cal L}_{01001} + {\cal L}_{10010}\right)\nonumber \\
&\quad+a_6\left[{\cal L}_{00101} + {\cal L}_{10100} + 2 ( {\cal L}_{00011} 
+ {\cal L}_{11000})  \right]
+a_7\left[{\cal L}_{11010} + {\cal L}_{01011} + 3 ( {\cal L}_{00011} 
+ {\cal L}_{11000}) \right]\nonumber \\
&\quad+a_8\,\zeta_2{\cal L}_{000}
+a_9\,\zeta_2\left({\cal L}_{001}+{\cal L}_{100}\right)
+a_{10}\,\zeta_3\,{\cal L}_{00}+a_{11}\,\zeta_2^2\,{\cal L}_0,
\end{align}
where each $a_i\in{\mathbb Q}$ is an undetermined rational numerical
coefficient, and we have suppressed the dependence of the SVHPLs on their
argument, ${\cal L}_w \equiv {\cal L}_w(z)$. Writing our ansatz for the 
constant~$C$ explicitly as
\begin{equation}
C=a_{12}\,\zeta_5+a_{13}\,\zeta_2\,\zeta_3\,,
\label{ansatz2}
\end{equation}

The final step in the bootstrap calculation is to constrain the parameters $a_1$ though $a_{13}$ using the information from the high-energy and the collinear limits reviewed in the last two sections. 

To use the high-energy limit one needs to consider the analytic continuation of the ansatz for $F(z)$ given by eq.~(\ref{ansatz1}) to the physical region of $2\to 2$ scattering, before taking the high-energy limit, $s\gg -t$. This process requires familiarity with the analytic properties of SVHPLs and is described in detail in section 5 of ref.~\cite{Almelid:2017qju}. It turns out that the  general ansatz of eq.~(\ref{ansatz1}) gives rise to multiple logarithmically-enhanced terms, $c_k L^k$ ranging all the way to $k=5$, which is the maximal power for weight-five polylogarithms. Setting to zero the real part of $c_k$ for any $k\geq 1$, and similarly the imaginary parts for any $k\geq 2$, leads to six constraints on the $a_i$ coefficients. This is then repeated for the three different forward limits, yielding a total of 8 independent linear relations between $a_i$:
\begin{align}\label{aconditions_regge}
\begin{split}
\!\!\!\!(a_1,\ldots,a_8) &= \left(0,\frac{a_{10}}{10},-\frac{a_{10}}{10}-\frac{a_{11}}{48},\frac{a_9}{2}-\frac{3a_{10}}{20}-\frac{a_{11}}{12},\frac{a_{10}}{10}+\frac{a_{11}}{48},\frac{7a_{10}}{20},-\frac{a_{10}}{4},-\frac{3a_{10}}{5}\right).
\end{split}
\end{align}

Finally, as shown in section 6 of ref.~\cite{Almelid:2017qju}, the collinear-limit constraint, namely the fact that the right hand side of eq.~(\ref{delcoll}) is a constant, independent of the colour degrees of freedom of the non-collinear partons, translates into the requirement that
\begin{align}
\Big[ \mathcal{F}(\rho_{1234}, \rho_{1432} ) - \mathcal{F}(\rho_{1243}, \rho_{1342} ) +  2 \mathcal{F}(\rho_{1324}, \rho_{1423} ) \Big]_{p_1 \parallel p_2} &= 0,
\notag \\
\Big[ \mathcal{F}(\rho_{1234}, \rho_{1432} ) + \mathcal{F}(\rho_{1243}, \rho_{1342} ) \Big]_{p_1 \parallel p_2} &= 8 C.
\label{Fcalcoll}
\end{align}
This, in turn,  implies the following relations between the coefficients $a_i$:
\begin{align}
(a_1,a_7,a_8,a_{11},a_{12},a_{13})=
\left(0,-\frac{a_{10}}{4},0,0,
3a_2-\frac{3a_3}{2}+a_4+\frac{9a_5}{2}-\frac{3a_6}{2}
+\frac{a_7}{2}, a_9\right).
\label{aconditions_coll2}
\end{align}
Upon implementing the full set of constraints, the
ans\"atze of eqs.~(\ref{ansatz1}) and~(\ref{ansatz2}) reduce to
\begin{equation}
F(z)=a_4\left({\cal L}_{10101}+2\zeta_2({\cal L}_{100}+{\cal L}_{001})
\right) {\rm~~~and~~~~}
C = a_4(\zeta_5+2\zeta_2\zeta_3)\,.
\label{ansatz4}
\end{equation}
Thus, both $F$ and $C$ have been uniquely determined by the bootstrap procedure,
up to an overall rational number $a_4$. Since $\Delta_n^{(3)}$ depends linearly on $F$ and $C$,
the form of the three-loop correction to the dipole formula is fixed completely by symmetries and physical constraints.
Comparing the expressions in eq.~\eqref{ansatz4} with the result of ref.~\cite{Almelid:2015jia} (quoted here in
eqs.~(\ref{Fdef}) and~(\ref{Cdef})), we see that the latter can be
reproduced by setting $a_4 = 1$.

\section{Conclusions}

Infrared singularities of massless scattering amplitudes in gauge theories are known in full to three-loop accuracy~\cite{Almelid:2015jia}. This is the first order in which corrections to the dipole sum occur. For three coloured partons this is a constant contribution, while for four or more it involves non-trivial kinematic dependence. 

Analysing special kinematic limits has proven very insightful, both as a way to learn about the QCD dynamics in these limits per se, and as a way to provide input for a bootstrap procedure. As done in ref.~\cite{Almelid:2017qju} and reviewed here,  at three loops the available information was sufficient to reproduce the result of the diagrammatic computation of the soft anomalous dimension up to a single overall numerical factor. 

In the high-energy limit there has been major progress in computing logarithmically-enhanced terms in $2\to 2$ amplitudes based on an order-by-order solution of the corresponding  rapidity evolution equations~\cite{Caron-Huot:2013fea,Caron-Huot:2017fxr,Caron-Huot:2017zfo}. In particular, at NLL we now know the soft anomalous dimension to all loop orders, and at NNLL we have an independent computation of both infrared-singular and finite terms at three loops for the real part of the amplitude.  The same methods can be applied to obtain explicit results at higher orders and infer the iterated structure of amplitudes in this limit.

We saw that understanding the space of functions in which the answer for the soft anomalous dimension should be written opens the way to use the general constraints from symmetries and the information from special kinematic limits to construct the answer. With these methods, the four-loop soft anomalous dimension may be within reach. Furthermore, other physical quantities, such as the single-emission radiative soft function may be amenable to similar methods. Finally, it is conceivable that bootstrap techniques could eventually be applied to multi-loop QCD amplitudes themselves. Of course, this still requires a significant advancement of our understanding of their properties. Their infrared singularities and special kinematic limits is a good place to start.

\begin{acknowledgments}
I would like to thank {\O}yvind Almelid, Claude Duhr, Simon Caron-Huot, Andrew McLeod,  Joscha Reichel, Leonardo Vernazza and Chris White for very stimulating and enjoyable collaboration on the recent work reported here. This work is supported by the STFC 
Consolidated Grant ``Particle Physics at the Higgs Centre''. 
\end{acknowledgments}

\bibliography{refs.bib}

\providecommand{\href}[2]{#2}\begingroup\raggedright\begin{thebibliography}{100}

\bibitem{Polyakov:1980ca}
A.~M. Polyakov, \emph{{Gauge Fields as Rings of Glue}},
  \href{http://dx.doi.org/10.1016/0550-3213(80)90507-6}{\emph{Nucl. Phys.} {\bf
  B164} (1980) 171--188}.

\bibitem{Arefeva:1980zd}
I.~Y. Arefeva, \emph{{Quantum contour field equations}},
  \href{http://dx.doi.org/10.1016/0370-2693(80)90529-8}{\emph{Phys. Lett.} {\bf
  B93} (1980) 347--353}.

\bibitem{Dotsenko:1979wb}
V.~S. Dotsenko and S.~N. Vergeles, \emph{{Renormalizability of Phase Factors in
  the Nonabelian Gauge Theory}},
  \href{http://dx.doi.org/10.1016/0550-3213(80)90103-0}{\emph{Nucl. Phys.} {\bf
  B169} (1980) 527}.

\bibitem{Brandt:1981kf}
R.~A. Brandt, F.~Neri and M.-a. Sato, \emph{{Renormalization of Loop Functions
  for All Loops}}, \href{http://dx.doi.org/10.1103/PhysRevD.24.879}{\emph{Phys.
  Rev.} {\bf D24} (1981) 879}.

\bibitem{Korchemsky:1985xj}
G.~P. Korchemsky and A.~V. Radyushkin, \emph{Loop space formalism and
  renormalization group for the infrared asymptotics of {QCD}},
  \href{http://dx.doi.org/10.1016/0370-2693(86)91439-5}{\emph{Phys. Lett.} {\bf
  B171} (1986) 459--467}.

\bibitem{Korchemsky:1985xu}
G.~Korchemsky and A.~Radyushkin, \emph{Infrared asymptotics of perturbative
  {QCD}: {R}enormalization properties of the wilson loops in higher orders of
  perturbation theory}, {\emph{Sov. J. Nucl. Phys.} {\bf 44} (1986) 877}.

\bibitem{Korchemsky:1987wg}
G.~Korchemsky and A.~Radyushkin, \emph{{Renormalization of the Wilson Loops
  Beyond the Leading Order}},
  \href{http://dx.doi.org/10.1016/0550-3213(87)90277-X}{\emph{Nucl. Phys.} {\bf
  B283} (1987) 342--364}.

\bibitem{Korchemskaya:1996je}
I.~Korchemskaya and G.~Korchemsky, \emph{{Evolution equation for gluon Regge
  trajectory}},
  \href{http://dx.doi.org/10.1016/0370-2693(96)01016-7}{\emph{Phys. Lett.} {\bf
  B387} (1996) 346--354}, [\href{https://arxiv.org/abs/hep-ph/9607229}{{\tt
  hep-ph/9607229}}].

\bibitem{Korchemskaya:1994qp}
I.~Korchemskaya and G.~Korchemsky, \emph{{High-energy scattering in QCD and
  cross singularities of Wilson loops}},
  \href{http://dx.doi.org/10.1016/0550-3213(94)00553-Q}{\emph{Nucl. Phys.} {\bf
  B437} (1995) 127--162}, [\href{https://arxiv.org/abs/hep-ph/9409446}{{\tt
  hep-ph/9409446}}].

\bibitem{Kuraev:1977fs}
E.~A. Kuraev, L.~N. Lipatov and V.~S. Fadin, \emph{{The Pomeranchuk Singularity
  in Nonabelian Gauge Theories}}, {\emph{Sov. Phys. JETP} {\bf 45} (1977)
  199--204}.

\bibitem{Balitsky:1978ic}
I.~I. Balitsky and L.~N. Lipatov, \emph{{The Pomeranchuk Singularity in Quantum
  Chromodynamics}}, {\emph{Sov. J. Nucl. Phys.} {\bf 28} (1978) 822--829}.

\bibitem{Lipatov:1985uk}
L.~N. Lipatov, \emph{{The Bare Pomeron in Quantum Chromodynamics}}, {\emph{Sov.
  Phys. JETP} {\bf 63} (1986) 904--912}.

\bibitem{Mueller:1993rr}
A.~H. Mueller, \emph{{Soft gluons in the infinite momentum wave function and
  the BFKL pomeron}},
  \href{http://dx.doi.org/10.1016/0550-3213(94)90116-3}{\emph{Nucl. Phys.} {\bf
  B415} (1994) 373--385}.

\bibitem{Mueller:1994jq}
A.~H. Mueller and B.~Patel, \emph{{Single and double BFKL pomeron exchange and
  a dipole picture of high-energy hard processes}},
  \href{http://dx.doi.org/10.1016/0550-3213(94)90284-4}{\emph{Nucl. Phys.} {\bf
  B425} (1994) 471--488}, [\href{https://arxiv.org/abs/hep-ph/9403256}{{\tt
  hep-ph/9403256}}].

\bibitem{Balitsky:1995ub}
I.~Balitsky, \emph{{Operator expansion for high-energy scattering}},
  \href{http://dx.doi.org/10.1016/0550-3213(95)00638-9}{\emph{Nucl. Phys.} {\bf
  B463} (1996) 99--160}, [\href{https://arxiv.org/abs/hep-ph/9509348}{{\tt
  hep-ph/9509348}}].

\bibitem{Balitsky:1998kc}
I.~Balitsky, \emph{{Factorization for high-energy scattering}},
  \href{http://dx.doi.org/10.1103/PhysRevLett.81.2024}{\emph{Phys. Rev. Lett.}
  {\bf 81} (1998) 2024--2027},
  [\href{https://arxiv.org/abs/hep-ph/9807434}{{\tt hep-ph/9807434}}].

\bibitem{Kovchegov:1999yj}
Y.~V. Kovchegov, \emph{{Small x F(2) structure function of a nucleus including
  multiple pomeron exchanges}},
  \href{http://dx.doi.org/10.1103/PhysRevD.60.034008}{\emph{Phys. Rev.} {\bf
  D60} (1999) 034008}, [\href{https://arxiv.org/abs/hep-ph/9901281}{{\tt
  hep-ph/9901281}}].

\bibitem{JalilianMarian:1996xn}
J.~Jalilian-Marian, A.~Kovner, L.~D. McLerran and H.~Weigert, \emph{{The
  Intrinsic glue distribution at very small x}},
  \href{http://dx.doi.org/10.1103/PhysRevD.55.5414}{\emph{Phys. Rev.} {\bf D55}
  (1997) 5414--5428}, [\href{https://arxiv.org/abs/hep-ph/9606337}{{\tt
  hep-ph/9606337}}].

\bibitem{JalilianMarian:1997gr}
J.~Jalilian-Marian, A.~Kovner, A.~Leonidov and H.~Weigert, \emph{{The Wilson
  renormalization group for low x physics: Towards the high density regime}},
  \href{http://dx.doi.org/10.1103/PhysRevD.59.014014}{\emph{Phys. Rev.} {\bf
  D59} (1998) 014014}, [\href{https://arxiv.org/abs/hep-ph/9706377}{{\tt
  hep-ph/9706377}}].

\bibitem{Iancu:2001ad}
E.~Iancu, A.~Leonidov and L.~D. McLerran, \emph{{The Renormalization group
  equation for the color glass condensate}},
  \href{http://dx.doi.org/10.1016/S0370-2693(01)00524-X}{\emph{Phys. Lett.}
  {\bf B510} (2001) 133--144},
  [\href{https://arxiv.org/abs/hep-ph/0102009}{{\tt hep-ph/0102009}}].

\bibitem{Caron-Huot:2013fea}
S.~Caron-Huot, \emph{{When does the gluon reggeize?}},
  \href{http://dx.doi.org/10.1007/JHEP05(2015)093}{\emph{JHEP} {\bf 05} (2015)
  093}, [\href{https://arxiv.org/abs/1309.6521}{{\tt 1309.6521}}].

\bibitem{Caron-Huot:2017fxr}
S.~Caron-Huot, E.~Gardi and L.~Vernazza, \emph{{Two-parton scattering in the
  high-energy limit}},
  \href{http://dx.doi.org/10.1007/JHEP06(2017)016}{\emph{JHEP} {\bf 06} (2017)
  016}, [\href{https://arxiv.org/abs/1701.05241}{{\tt 1701.05241}}].

\bibitem{Caron-Huot:2017zfo}
S.~Caron-Huot, E.~Gardi, J.~Reichel and L.~Vernazza, \emph{{Infrared
  singularities of QCD scattering amplitudes in the Regge limit to all
  orders}},  \href{https://arxiv.org/abs/1711.04850}{{\tt 1711.04850}}.

\bibitem{Mueller:1979ih}
A.~H. Mueller, \emph{On the asymptotic behavior of the {S}udakov form-factor},
  \href{http://dx.doi.org/10.1103/PhysRevD.20.2037}{\emph{Phys. Rev.} {\bf D20}
  (1979) 2037}.

\bibitem{Collins:1980ih}
J.~C. Collins, \emph{Algorithm to compute corrections to the {S}udakov
  form-factor}, \href{http://dx.doi.org/10.1103/PhysRevD.22.1478}{\emph{Phys.
  Rev.} {\bf D22} (1980) 1478}.

\bibitem{Sen:1981sd}
A.~Sen, \emph{{Asymptotic Behavior of the Sudakov Form-Factor in QCD}},
  \href{http://dx.doi.org/10.1103/PhysRevD.24.3281}{\emph{Phys. Rev.} {\bf D24}
  (1981) 3281}.

\bibitem{Sen:1982bt}
A.~Sen, \emph{{Asymptotic Behavior of the Wide Angle On-Shell Quark Scattering
  Amplitudes in Nonabelian Gauge Theories}},
  \href{http://dx.doi.org/10.1103/PhysRevD.28.860}{\emph{Phys. Rev.} {\bf D28}
  (1983) 860}.

\bibitem{Gatheral:1983cz}
J.~G.~M. Gatheral, \emph{{Exponentiation of eikonal cross-sections in
  nonabelian gauge theories}},
  \href{http://dx.doi.org/10.1016/0370-2693(83)90112-0}{\emph{Phys. Lett.} {\bf
  B133} (1983) 90}.

\bibitem{Frenkel:1984pz}
J.~Frenkel and J.~C. Taylor, \emph{{Nonabelian eikonal exponentiation}},
  \href{http://dx.doi.org/10.1016/0550-3213(84)90294-3}{\emph{Nucl. Phys.} {\bf
  B246} (1984) 231}.

\bibitem{Sterman:1981jc}
G.~F. Sterman, \emph{Infrared divergences in perturbative {QCD}. (talk)},
  \href{http://dx.doi.org/10.1063/1.33099}{\emph{AIP Conf. Proc.} 22--40}.

\bibitem{Magnea:1990zb}
L.~Magnea and G.~F. Sterman, \emph{{Analytic continuation of the Sudakov
  form-factor in QCD}},
  \href{http://dx.doi.org/10.1103/PhysRevD.42.4222}{\emph{Phys. Rev.} {\bf D42}
  (1990) 4222--4227}.

\bibitem{Korchemsky:1993hr}
G.~P. Korchemsky, \emph{{On Near forward high-energy scattering in QCD}},
  \href{http://dx.doi.org/10.1016/0370-2693(94)90040-X}{\emph{Phys. Lett.} {\bf
  B325} (1994) 459--466}, [\href{https://arxiv.org/abs/hep-ph/9311294}{{\tt
  hep-ph/9311294}}].

\bibitem{Catani:1996vz}
S.~Catani and M.~H. Seymour, \emph{{A general algorithm for calculating jet
  cross sections in NLO QCD}},
  \href{http://dx.doi.org/10.1016/S0550-3213(96)00589-5}{\emph{Nucl. Phys.}
  {\bf B485} (1997) 291--419},
  [\href{https://arxiv.org/abs/hep-ph/9605323}{{\tt hep-ph/9605323}}].

\bibitem{Catani:1998bh}
S.~Catani, \emph{{The Singular behavior of QCD amplitudes at two loop order}},
  \href{http://dx.doi.org/10.1016/S0370-2693(98)00332-3}{\emph{Phys. Lett.}
  {\bf B427} (1998) 161--171},
  [\href{https://arxiv.org/abs/hep-ph/9802439}{{\tt hep-ph/9802439}}].

\bibitem{Sterman:2002qn}
G.~F. Sterman and M.~E. Tejeda-Yeomans, \emph{{Multi-loop amplitudes and
  resummation}},
  \href{http://dx.doi.org/10.1016/S0370-2693(02)03100-3}{\emph{Phys. Lett.}
  {\bf B552} (2003) 48--56}, [\href{https://arxiv.org/abs/hep-ph/0210130}{{\tt
  hep-ph/0210130}}].

\bibitem{Dixon:2008gr}
L.~J. Dixon, L.~Magnea and G.~Sterman, \emph{{Universal structure of subleading
  infrared poles in gauge theory amplitudes}},
  \href{http://dx.doi.org/10.1088/1126-6708/2008/08/022}{\emph{JHEP} {\bf 08}
  (2008) 022}, [\href{https://arxiv.org/abs/0805.3515}{{\tt 0805.3515}}].

\bibitem{Kidonakis:1998nf}
N.~Kidonakis, G.~Oderda and G.~F. Sterman, \emph{{Evolution of color exchange
  in {QCD} hard scattering}},
  \href{http://dx.doi.org/10.1016/S0550-3213(98)00441-6}{\emph{Nucl. Phys.}
  {\bf B531} (1998) 365--402},
  [\href{https://arxiv.org/abs/hep-ph/9803241}{{\tt hep-ph/9803241}}].

\bibitem{Bonciani:2003nt}
R.~Bonciani, S.~Catani, M.~L. Mangano and P.~Nason, \emph{{Sudakov resummation
  of multiparton QCD cross-sections}},
  \href{http://dx.doi.org/10.1016/j.physletb.2003.09.068}{\emph{Phys. Lett.}
  {\bf B575} (2003) 268--278},
  [\href{https://arxiv.org/abs/hep-ph/0307035}{{\tt hep-ph/0307035}}].

\bibitem{Dokshitzer:2005ig}
{\relax Yu}.~L. Dokshitzer and G.~Marchesini, \emph{{Soft gluons at large
  angles in hadron collisions}},
  \href{http://dx.doi.org/10.1088/1126-6708/2006/01/007}{\emph{JHEP} {\bf 01}
  (2006) 007}, [\href{https://arxiv.org/abs/hep-ph/0509078}{{\tt
  hep-ph/0509078}}].

\bibitem{Aybat:2006mz}
S.~M. Aybat, L.~J. Dixon and G.~F. Sterman, \emph{{The two-loop soft anomalous
  dimension matrix and resummation at next-to-next-to leading pole}},
  \href{http://dx.doi.org/10.1103/PhysRevD.74.074004}{\emph{Phys. Rev.} {\bf
  D74} (2006) 074004}, [\href{https://arxiv.org/abs/hep-ph/0607309}{{\tt
  hep-ph/0607309}}].

\bibitem{Gardi:2009qi}
E.~Gardi and L.~Magnea, \emph{{Factorization constraints for soft anomalous
  dimensions in QCD scattering amplitudes}},
  \href{http://dx.doi.org/10.1088/1126-6708/2009/03/079}{\emph{JHEP} {\bf 0903}
  (2009) 079}, [\href{https://arxiv.org/abs/0901.1091}{{\tt 0901.1091}}].

\bibitem{Becher:2009cu}
T.~Becher and M.~Neubert, \emph{{Infrared singularities of scattering
  amplitudes in perturbative QCD}},
  \href{http://dx.doi.org/10.1103/PhysRevLett.102.162001}{\emph{Phys. Rev.
  Lett.} {\bf 102} (2009) 162001}, [\href{https://arxiv.org/abs/0901.0722}{{\tt
  0901.0722}}].

\bibitem{Becher:2009qa}
T.~Becher and M.~Neubert, \emph{{On the Structure of Infrared Singularities of
  Gauge-Theory Amplitudes}},
  \href{http://dx.doi.org/10.1088/1126-6708/2009/06/081}{\emph{JHEP} {\bf 06}
  (2009) 081}, [\href{https://arxiv.org/abs/0903.1126}{{\tt 0903.1126}}].

\bibitem{Gardi:2009zv}
E.~Gardi and L.~Magnea, \emph{{Infrared singularities in QCD amplitudes}},
  \href{http://dx.doi.org/10.1393/ncc/i2010-10528-x}{\emph{Nuovo Cim.} {\bf
  032C} (2009) 137--157}, [\href{https://arxiv.org/abs/0908.3273}{{\tt
  0908.3273}}].

\bibitem{Dixon:2009gx}
L.~J. Dixon, \emph{{Matter Dependence of the Three-Loop Soft Anomalous
  Dimension Matrix}},
  \href{http://dx.doi.org/10.1103/PhysRevD.79.091501}{\emph{Phys. Rev.} {\bf
  D79} (2009) 091501}, [\href{https://arxiv.org/abs/0901.3414}{{\tt
  0901.3414}}].

\bibitem{Dixon:2009ur}
L.~J. Dixon, E.~Gardi and L.~Magnea, \emph{{On soft singularities at three
  loops and beyond}},
  \href{http://dx.doi.org/10.1007/JHEP02(2010)081}{\emph{JHEP} {\bf 02} (2010)
  081}, [\href{https://arxiv.org/abs/0910.3653}{{\tt 0910.3653}}].

\bibitem{Bret:2011xm}
V.~Del~Duca, C.~Duhr, E.~Gardi, L.~Magnea and C.~D. White, \emph{{An infrared
  approach to Reggeization}},
  \href{http://dx.doi.org/10.1103/PhysRevD.85.071104}{\emph{Phys.Rev.} {\bf
  D85} (2012) 071104}, [\href{https://arxiv.org/abs/1108.5947}{{\tt
  1108.5947}}].

\bibitem{DelDuca:2011ae}
V.~Del~Duca, C.~Duhr, E.~Gardi, L.~Magnea and C.~D. White, \emph{{The Infrared
  structure of gauge theory amplitudes in the high-energy limit}},
  \href{http://dx.doi.org/10.1007/JHEP12(2011)021}{\emph{JHEP} {\bf 1112}
  (2011) 021}, [\href{https://arxiv.org/abs/1109.3581}{{\tt 1109.3581}}].

\bibitem{Ahrens:2012qz}
V.~Ahrens, M.~Neubert and L.~Vernazza, \emph{{Structure of Infrared
  Singularities of Gauge-Theory Amplitudes at Three and Four Loops}},
  \href{http://dx.doi.org/10.1007/JHEP09(2012)138}{\emph{JHEP} {\bf 1209}
  (2012) 138}, [\href{https://arxiv.org/abs/1208.4847}{{\tt 1208.4847}}].

\bibitem{Naculich:2013xa}
S.~G. Naculich, H.~Nastase and H.~J. Schnitzer, \emph{{All-loop
  infrared-divergent behavior of most-subleading-color gauge-theory
  amplitudes}}, \href{http://dx.doi.org/10.1007/JHEP04(2013)114}{\emph{JHEP}
  {\bf 1304} (2013) 114}, [\href{https://arxiv.org/abs/1301.2234}{{\tt
  1301.2234}}].

\bibitem{Erdogan:2014gha}
O.~Erdo{\u g}an and G.~Sterman, \emph{{Ultraviolet divergences and
  factorization for coordinate-space amplitudes}},
  \href{http://dx.doi.org/10.1103/PhysRevD.91.065033}{\emph{Phys. Rev.} {\bf
  D91} (2015) 065033}, [\href{https://arxiv.org/abs/1411.4588}{{\tt
  1411.4588}}].

\bibitem{Gehrmann:2010ue}
T.~Gehrmann, E.~W.~N. Glover, T.~Huber, N.~Ikizlerli and C.~Studerus,
  \emph{{Calculation of the quark and gluon form factors to three loops in
  QCD}}, \href{http://dx.doi.org/10.1007/JHEP06(2010)094}{\emph{JHEP} {\bf 06}
  (2010) 094}, [\href{https://arxiv.org/abs/1004.3653}{{\tt 1004.3653}}].

\bibitem{Kidonakis:2009ev}
N.~Kidonakis, \emph{{Two-loop soft anomalous dimensions and NNLL resummation
  for heavy quark production}},
  \href{http://dx.doi.org/10.1103/PhysRevLett.102.232003}{\emph{Phys. Rev.
  Lett.} {\bf 102} (2009) 232003}, [\href{https://arxiv.org/abs/0903.2561}{{\tt
  0903.2561}}].

\bibitem{Mitov:2009sv}
A.~Mitov, G.~Sterman and I.~Sung, \emph{{The Massive Soft Anomalous Dimension
  Matrix at Two Loops}},
  \href{http://dx.doi.org/10.1103/PhysRevD.79.094015}{\emph{Phys. Rev.} {\bf
  D79} (2009) 094015}, [\href{https://arxiv.org/abs/0903.3241}{{\tt
  0903.3241}}].

\bibitem{Becher:2009kw}
T.~Becher and M.~Neubert, \emph{{Infrared singularities of QCD amplitudes with
  massive partons}},
  \href{http://dx.doi.org/10.1103/PhysRevD.79.125004}{\emph{Phys. Rev.} {\bf
  D79} (2009) 125004}, [\href{https://arxiv.org/abs/0904.1021}{{\tt
  0904.1021}}].

\bibitem{Beneke:2009rj}
M.~Beneke, P.~Falgari and C.~Schwinn, \emph{{Soft radiation in heavy-particle
  pair production: all- order colour structure and two-loop anomalous
  dimension}},
  \href{http://dx.doi.org/10.1016/j.nuclphysb.2009.11.004}{\emph{Nucl. Phys.}
  {\bf B828} (2010) 69--101}, [\href{https://arxiv.org/abs/0907.1443}{{\tt
  0907.1443}}].

\bibitem{Czakon:2009zw}
M.~Czakon, A.~Mitov and G.~F. Sterman, \emph{{Threshold Resummation for
  Top-Pair Hadroproduction to Next-to-Next-to-Leading Log}},
  \href{http://dx.doi.org/10.1103/PhysRevD.80.074017}{\emph{Phys. Rev.} {\bf
  D80} (2009) 074017}, [\href{https://arxiv.org/abs/0907.1790}{{\tt
  0907.1790}}].

\bibitem{Ferroglia:2009ii}
A.~Ferroglia, M.~Neubert, B.~D. Pecjak and L.~L. Yang, \emph{{Two-loop
  divergences of massive scattering amplitudes in non-abelian gauge theories}},
  \href{http://dx.doi.org/10.1088/1126-6708/2009/11/062}{\emph{JHEP} {\bf 11}
  (2009) 062}, [\href{https://arxiv.org/abs/0908.3676}{{\tt 0908.3676}}].

\bibitem{Chiu:2009mg}
J.-y. Chiu, A.~Fuhrer, R.~Kelley and A.~V. Manohar, \emph{{Factorization
  Structure of Gauge Theory Amplitudes and Application to Hard Scattering
  Processes at the LHC}},
  \href{http://dx.doi.org/10.1103/PhysRevD.80.094013}{\emph{Phys. Rev.} {\bf
  D80} (2009) 094013}, [\href{https://arxiv.org/abs/0909.0012}{{\tt
  0909.0012}}].

\bibitem{Mitov:2010rp}
A.~Mitov, G.~Sterman and I.~Sung, \emph{{Diagrammatic Exponentiation for
  Products of Wilson Lines}},
  \href{http://dx.doi.org/10.1103/PhysRevD.82.096010}{\emph{Phys. Rev.} {\bf
  D82} (2010) 096010}, [\href{https://arxiv.org/abs/1008.0099}{{\tt
  1008.0099}}].

\bibitem{Mitov:2010xw}
A.~Mitov, G.~F. Sterman and I.~Sung, \emph{{Computation of the Soft Anomalous
  Dimension Matrix in Coordinate Space}},
  \href{http://dx.doi.org/10.1103/PhysRevD.82.034020}{\emph{Phys. Rev.} {\bf
  D82} (2010) 034020}, [\href{https://arxiv.org/abs/1005.4646}{{\tt
  1005.4646}}].

\bibitem{Gardi:2013saa}
E.~Gardi, \emph{{From Webs to Polylogarithms}},
  \href{http://dx.doi.org/10.1007/JHEP04(2014)044}{\emph{JHEP} {\bf 1404}
  (2014) 044}, [\href{https://arxiv.org/abs/1310.5268}{{\tt 1310.5268}}].

\bibitem{Falcioni:2014pka}
G.~Falcioni, E.~Gardi, M.~Harley, L.~Magnea and C.~D. White, \emph{{Multiple
  Gluon Exchange Webs}},
  \href{http://dx.doi.org/10.1007/JHEP10(2014)010}{\emph{JHEP} {\bf 10} (2014)
  10}, [\href{https://arxiv.org/abs/1407.3477}{{\tt 1407.3477}}].

\bibitem{Henn:2013fah}
J.~M. Henn, A.~V. Smirnov and V.~A. Smirnov, \emph{{Analytic results for planar
  three-loop four-point integrals from a Knizhnik-Zamolodchikov equation}},
  \href{http://dx.doi.org/10.1007/JHEP07(2013)128}{\emph{JHEP} {\bf 07} (2013)
  128}, [\href{https://arxiv.org/abs/1306.2799}{{\tt 1306.2799}}].

\bibitem{Dukes:2013gea}
M.~Dukes, E.~Gardi, H.~McAslan, D.~J. Scott and C.~D. White, \emph{{Webs and
  Posets}}, \href{http://dx.doi.org/10.1007/JHEP01(2014)024}{\emph{JHEP} {\bf
  01} (2014) 024}, [\href{https://arxiv.org/abs/1310.3127}{{\tt 1310.3127}}].

\bibitem{Dukes:2013wa}
M.~Dukes, E.~Gardi, E.~Steingrimsson and C.~D. White, \emph{{Web worlds,
  web-colouring matrices, and web-mixing matrices}},
  \href{http://dx.doi.org/10.1016/j.jcta.2013.02.001}{\emph{J. Comb. Theory
  Ser. A} {\bf 120} (2013) 1012--1037},
  [\href{https://arxiv.org/abs/1301.6576}{{\tt 1301.6576}}].

\bibitem{Gardi:2010rn}
E.~Gardi, E.~Laenen, G.~Stavenga and C.~D. White, \emph{{Webs in multiparton
  scattering using the replica trick}},
  \href{http://dx.doi.org/10.1007/JHEP11(2010)155}{\emph{JHEP} {\bf 1011}
  (2010) 155}, [\href{https://arxiv.org/abs/1008.0098}{{\tt 1008.0098}}].

\bibitem{Gardi:2011wa}
E.~Gardi and C.~D. White, \emph{{General properties of multiparton webs: Proofs
  from combinatorics}},
  \href{http://dx.doi.org/10.1007/JHEP03(2011)079}{\emph{JHEP} {\bf 1103}
  (2011) 079}, [\href{https://arxiv.org/abs/1102.0756}{{\tt 1102.0756}}].

\bibitem{Gardi:2011yz}
E.~Gardi, J.~M. Smillie and C.~D. White, \emph{{On the renormalization of
  multiparton webs}},
  \href{http://dx.doi.org/10.1007/JHEP09(2011)114}{\emph{JHEP} {\bf 1109}
  (2011) 114}, [\href{https://arxiv.org/abs/1108.1357}{{\tt 1108.1357}}].

\bibitem{Gardi:2013ita}
E.~Gardi, J.~M. Smillie and C.~D. White, \emph{{The Non-Abelian Exponentiation
  theorem for multiple Wilson lines}},
  \href{http://dx.doi.org/10.1007/JHEP06(2013)088}{\emph{JHEP} {\bf 06} (2013)
  088}, [\href{https://arxiv.org/abs/1304.7040}{{\tt 1304.7040}}].

\bibitem{Laenen:2008gt}
E.~Laenen, G.~Stavenga and C.~D. White, \emph{{Path integral approach to
  eikonal and next-to-eikonal exponentiation}},
  \href{http://dx.doi.org/10.1088/1126-6708/2009/03/054}{\emph{JHEP} {\bf 03}
  (2009) 054}, [\href{https://arxiv.org/abs/0811.2067}{{\tt 0811.2067}}].

\bibitem{Vladimirov:2015fea}
A.~A. Vladimirov, \emph{{Exponentiation for products of Wilson lines within the
  generating function approach}},
  \href{http://dx.doi.org/10.1007/JHEP06(2015)120}{\emph{JHEP} {\bf 06} (2015)
  120}, [\href{https://arxiv.org/abs/1501.03316}{{\tt 1501.03316}}].

\bibitem{Almelid:2015jia}
{\O}.~Almelid, C.~Duhr and E.~Gardi, \emph{{Three-loop corrections to the soft
  anomalous dimension in multileg scattering}},
  \href{http://dx.doi.org/10.1103/PhysRevLett.117.172002}{\emph{Phys. Rev.
  Lett.} {\bf 117} (2016) 172002},
  [\href{https://arxiv.org/abs/1507.00047}{{\tt 1507.00047}}].

\bibitem{Gardi:2016ttq}
E.~Gardi, {\O}.~Almelid and C.~Duhr, \emph{{Long-distance singularities in
  multi-leg scattering amplitudes}}, {\emph{PoS} {\bf LL2016} (2016) 058},
  [\href{https://arxiv.org/abs/1606.05697}{{\tt 1606.05697}}].

\bibitem{Almelid:2017qju}
{\O}.~Almelid, C.~Duhr, E.~Gardi, A.~McLeod and C.~D. White,
  \emph{{Bootstrapping the QCD soft anomalous dimension}},
  \href{http://dx.doi.org/10.1007/JHEP09(2017)073}{\emph{JHEP} {\bf 09} (2017)
  073}, [\href{https://arxiv.org/abs/1706.10162}{{\tt 1706.10162}}].

\bibitem{Moch:2004pa}
S.~Moch, J.~Vermaseren and A.~Vogt, \emph{{The Three loop splitting functions
  in QCD: The Nonsinglet case}},
  \href{http://dx.doi.org/10.1016/j.nuclphysb.2004.03.030}{\emph{Nucl. Phys.}
  {\bf B688} (2004) 101--134},
  [\href{https://arxiv.org/abs/hep-ph/0403192}{{\tt hep-ph/0403192}}].

\bibitem{Moch:2005tm}
S.~Moch, J.~A.~M. Vermaseren and A.~Vogt, \emph{{Three-loop results for quark
  and gluon form-factors}},
  \href{http://dx.doi.org/10.1016/j.physletb.2005.08.067}{\emph{Phys. Lett.}
  {\bf B625} (2005) 245--252},
  [\href{https://arxiv.org/abs/hep-ph/0508055}{{\tt hep-ph/0508055}}].

\bibitem{Boels:2017skl}
R.~H. Boels, T.~Huber and G.~Yang, \emph{{The four-loop non-planar cusp
  anomalous dimension in N = 4 SYM}},
  \href{https://arxiv.org/abs/1705.03444}{{\tt 1705.03444}}.

\bibitem{Boels:2017ftb}
R.~H. Boels, T.~Huber and G.~Yang, \emph{{The Sudakov form factor at four loops
  in maximal super Yang-Mills theory}},
  \href{https://arxiv.org/abs/1711.08449}{{\tt 1711.08449}}.

\bibitem{Moch:2017uml}
S.~Moch, B.~Ruijl, T.~Ueda, J.~A.~M. Vermaseren and A.~Vogt, \emph{{Four-Loop
  Non-Singlet Splitting Functions in the Planar Limit and Beyond}},
  \href{http://dx.doi.org/10.1007/JHEP10(2017)041}{\emph{JHEP} {\bf 10} (2017)
  041}, [\href{https://arxiv.org/abs/1707.08315}{{\tt 1707.08315}}].

\bibitem{Grozin:2017css}
A.~Grozin, J.~Henn and M.~Stahlhofen, \emph{{On the Casimir scaling violation
  in the cusp anomalous dimension at small angle}},
  \href{http://dx.doi.org/10.1007/JHEP10(2017)052}{\emph{JHEP} {\bf 10} (2017)
  052}, [\href{https://arxiv.org/abs/1708.01221}{{\tt 1708.01221}}].

\bibitem{Brown:2004}
F.~C.~S. Brown, \emph{{Single-valued multiple polylogarithms in one variable}},
  \href{http://dx.doi.org/10.1016/j.crma.2004.02.001}{\emph{C. R. Acad. Sci.
  Paris} {\bf Ser. I} (2004) 338}.

\bibitem{Remiddi:1999ew}
E.~Remiddi and J.~Vermaseren, \emph{{Harmonic polylogarithms}},
  \href{http://dx.doi.org/10.1142/S0217751X00000367}{\emph{Int.J.Mod.Phys.}
  {\bf A15} (2000) 725--754}, [\href{https://arxiv.org/abs/hep-ph/9905237}{{\tt
  hep-ph/9905237}}].

\bibitem{Dixon:2012yy}
L.~J. Dixon, C.~Duhr and J.~Pennington, \emph{{Single-valued harmonic
  polylogarithms and the multi-Regge limit}},
  \href{http://dx.doi.org/10.1007/JHEP10(2012)074}{\emph{JHEP} {\bf 10} (2012)
  074}, [\href{https://arxiv.org/abs/1207.0186}{{\tt 1207.0186}}].

\bibitem{Brown:2009qja}
F.~C.~S. Brown, \emph{{Multiple zeta values and periods of moduli spaces M 0 ,n
  ( R )}}, {\emph{Annales Sci. Ecole Norm. Sup.} {\bf 42} (2009) 371},
  [\href{https://arxiv.org/abs/math/0606419}{{\tt math/0606419}}].

\bibitem{Sotiropoulos:1993rd}
M.~G. Sotiropoulos and G.~F. Sterman, \emph{{Color exchange in near forward
  hard elastic scattering}},
  \href{http://dx.doi.org/10.1016/0550-3213(94)90357-3}{\emph{Nucl. Phys.} {\bf
  B419} (1994) 59--76}, [\href{https://arxiv.org/abs/hep-ph/9310279}{{\tt
  hep-ph/9310279}}].

\bibitem{DelDuca:2001gu}
V.~Del~Duca and E.~W.~N. Glover, \emph{{The High-energy limit of QCD at two
  loops}}, \href{http://dx.doi.org/10.1088/1126-6708/2001/10/035}{\emph{JHEP}
  {\bf 10} (2001) 035}, [\href{https://arxiv.org/abs/hep-ph/0109028}{{\tt
  hep-ph/0109028}}].

\bibitem{DelDuca:2013ara}
V.~Del~Duca, G.~Falcioni, L.~Magnea and L.~Vernazza, \emph{{High-energy QCD
  amplitudes at two loops and beyond}},
  \href{http://dx.doi.org/10.1016/j.physletb.2014.03.033}{\emph{Phys. Lett.}
  {\bf B732} (2014) 233--240}, [\href{https://arxiv.org/abs/1311.0304}{{\tt
  1311.0304}}].

\bibitem{DelDuca:2014cya}
V.~Del~Duca, G.~Falcioni, L.~Magnea and L.~Vernazza, \emph{{Analyzing
  high-energy factorization beyond next-to-leading logarithmic accuracy}},
  \href{http://dx.doi.org/10.1007/JHEP02(2015)029}{\emph{JHEP} {\bf 02} (2015)
  029}, [\href{https://arxiv.org/abs/1409.8330}{{\tt 1409.8330}}].

\bibitem{Fadin:2016wso}
V.~S. Fadin, \emph{{Particularities of the NNLLA BFKL}},
  \href{http://dx.doi.org/10.1063/1.4977159}{\emph{AIP Conf. Proc.} {\bf 1819}
  (2017) 060003}, [\href{https://arxiv.org/abs/1612.04481}{{\tt 1612.04481}}].

\bibitem{Fadin:2017nka}
V.~S. Fadin and L.~N. Lipatov, \emph{{Reggeon cuts in QCD amplitudes with
  negative signature}},  \href{https://arxiv.org/abs/1712.09805}{{\tt
  1712.09805}}.

\bibitem{Henn:2016jdu}
J.~M. Henn and B.~Mistlberger, \emph{{Four-Gluon Scattering at Three Loops,
  Infrared Structure, and the Regge Limit}},
  \href{http://dx.doi.org/10.1103/PhysRevLett.117.171601}{\emph{Phys. Rev.
  Lett.} {\bf 117} (2016) 171601},
  [\href{https://arxiv.org/abs/1608.00850}{{\tt 1608.00850}}].

\bibitem{DelDuca:2013dsa}
V.~Del~Duca, G.~Falcioni, L.~Magnea and L.~Vernazza, \emph{{Beyond Reggeization
  for two- and three-loop QCD amplitudes}}, {\emph{PoS} {\bf RADCOR2013} (2013)
  046}, [\href{https://arxiv.org/abs/1312.5098}{{\tt 1312.5098}}].

\bibitem{Catani:2011st}
S.~Catani, D.~de~Florian and G.~Rodrigo, \emph{{Space-like (versus time-like)
  collinear limits in QCD: Is factorization violated?}},
  \href{http://dx.doi.org/10.1007/JHEP07(2012)026}{\emph{JHEP} {\bf 1207}
  (2012) 026}, [\href{https://arxiv.org/abs/1112.4405}{{\tt 1112.4405}}].

\bibitem{Forshaw:2012bi}
J.~R. Forshaw, M.~H. Seymour and A.~Siodmok, \emph{{On the Breaking of
  Collinear Factorization in QCD}},
  \href{http://dx.doi.org/10.1007/JHEP11(2012)066}{\emph{JHEP} {\bf 1211}
  (2012) 066}, [\href{https://arxiv.org/abs/1206.6363}{{\tt 1206.6363}}].

\bibitem{Bern:1999ry}
Z.~Bern, V.~Del~Duca, W.~B. Kilgore and C.~R. Schmidt, \emph{{The infrared
  behavior of one loop QCD amplitudes at next-to-next-to leading order}},
  \href{http://dx.doi.org/10.1103/PhysRevD.60.116001}{\emph{Phys. Rev.} {\bf
  D60} (1999) 116001}, [\href{https://arxiv.org/abs/hep-ph/9903516}{{\tt
  hep-ph/9903516}}].

\bibitem{Kosower:1999xi}
D.~A. Kosower, \emph{{All order collinear behavior in gauge theories}},
  \href{http://dx.doi.org/10.1016/S0550-3213(99)00251-5}{\emph{Nucl. Phys.}
  {\bf B552} (1999) 319--336},
  [\href{https://arxiv.org/abs/hep-ph/9901201}{{\tt hep-ph/9901201}}].

\bibitem{Feige:2014wja}
I.~Feige and M.~D. Schwartz, \emph{{Hard-Soft-Collinear Factorization to All
  Orders}}, \href{http://dx.doi.org/10.1103/PhysRevD.90.105020}{\emph{Phys.
  Rev.} {\bf D90} (2014) 105020}, [\href{https://arxiv.org/abs/1403.6472}{{\tt
  1403.6472}}].

\bibitem{Catani:2003vu}
S.~Catani, D.~de~Florian and G.~Rodrigo, \emph{{The Triple collinear limit of
  one loop QCD amplitudes}},
  \href{http://dx.doi.org/10.1016/j.physletb.2004.02.039}{\emph{Phys. Lett.}
  {\bf B586} (2004) 323--331},
  [\href{https://arxiv.org/abs/hep-ph/0312067}{{\tt hep-ph/0312067}}].

\bibitem{AlmelidPhD}
{\O}.~Almelid, \emph{The Three-Loop Soft Anomalous Dimension of Massless
  Multileg Scattering}.
\newblock PhD thesis, University of Edinburgh, 2016.

\bibitem{longinprep}
{\O}.~Almelid, C.~Duhr and E.~Gardi, \emph{Calculation of the soft anomalous
  dimension in multileg massless scattering at three loops},
  \href{https://arxiv.org/abs/In preparation}{{\tt In preparation}}.

\end{thebibliography}\endgroup

\end{document}